\documentclass[a4paper, 10pt, twoside]{article}

\usepackage{geometry}
\usepackage{algorithm}
\usepackage{algpseudocode}

\usepackage{graphicx}
\graphicspath{{./Figs/}}
\usepackage{subcaption}
\usepackage{multirow, booktabs, array, siunitx}
\usepackage{textcomp}
\usepackage[square,comma,sort&compress]{natbib}
\usepackage{times}
\usepackage{amsmath}
\usepackage{amsthm}
\usepackage{amsfonts}
\usepackage{tikz}
\usetikzlibrary{shapes.geometric, arrows}
\usepackage[toc,page]{appendix}

\usepackage[colorlinks,allcolors=blue]{hyperref}

\usepackage[toc,page]{appendix}
\usepackage{sectsty}

\usepackage{cleveref}
\usepackage{appendix}

\usepackage{tabularx,booktabs}
\newcolumntype{Y}{>{\centering\arraybackslash}X}

\usepackage{adjustbox}
\usepackage{xcolor}
\usepackage[normalem]{ulem}
\usepackage{marginnote}

\newcommand{\E}{\mathbf{E}}
\renewcommand{\d}{\mathrm{d}}

\theoremstyle{theorem}

\theoremstyle{definition}

\theoremstyle{definition}

\theoremstyle{theorem}

\title{Hedging with Linear Regressions and Neural Networks\footnotetext[0]{We thank Matthias B\"uchner, Agostino Capponi, Philipp D\"orsek, Ale\v{s} \v{C}ern\'{y}, Jean-Pierre Fouque, Camilo Garcia, Lukas Gonon, Harald Oberhauser, Philipp Illeditsch, Antoine Jacquier, Johannes Muhle-Karbe, Peter Spoida, Josef Teichmann, and James Wolter for helpful discussions on the subject matter of this paper.  We are grateful to Deutsche B\"orse, in particular, Peter Spoida, for providing us with Euro Stoxx~50 options and futures tick data. We are indebted to two anonymous referees and an associate editor for several very insightful comments that improved the paper. The code to reproduce results in this paper can be found at  \url{https://github.com/weiguanwang/Hedging_Neural_Networks}.}}

\author{Johannes Ruf\thanks{Department of Mathematics, London School of Economics and Political Science. Email: j.ruf@lse.ac.uk} \and Weiguan Wang\thanks{School of Economics, Shanghai University. Email: weiguanwang@outlook.com}}

\begin{document}

\maketitle

\begin{abstract}
We study  neural networks as nonparametric estimation tools for the hedging of  options. To this end,  we design a network, named \textit{HedgeNet}, that directly outputs a hedging strategy. This network is trained to minimise the hedging error instead of the pricing error. Applied to end-of-day and tick prices of S\&P~500 and Euro Stoxx~50 options, the network is able to reduce the mean squared hedging error of the Black-Scholes benchmark significantly.  However, a similar benefit arises by  simple linear regressions  that incorporate the leverage effect. 
\end{abstract}
\textbf{Keywords:} Benchmarking; Black-Scholes; Hedging error; Information leakage; Leverage effect; Statistical hedging

\section{Introduction}
Beginning with \cite{hutchinson1994nonparametric} and \cite{malliaris1993neural}, \textit{artificial neural networks} (ANNs) are being proposed as a nonparametric tool for the risk management of options. Since then  about 150 papers have been published that apply ANNs to price and hedge options; see Section~\ref{sec:network} for several pointers to this literature.
We show that for the estimation of the optimal hedging ratio ANNs do \textit{not} outperform simple linear regressions that use only standard option sensitivities.

We study a specific and well defined risk management application, namely the reduction of variance of the hedging error in daily options' trading. 
More precisely, we consider a one-period model and imagine an operator who is short an option (or a cross section of options). The mark-to-market accounting convention requires  a good control of the hedging error for short periods, even when considering long-dated options.  To reduce the variance of her portfolio the operator is allowed 
to buy or sell the underlying. Today, she sells the option, say at price $C_0$. She is now allowed to buy $\delta$ shares of the underlying at price $S_0$ and $C_0 - \delta S_0$ units of the risk-free asset. Then today's portfolio value equals $V_0 = 0$. Tomorrow, her portfolio has value
\begin{align} \label{eq:191020.1}
	V_1^\delta =  \delta S_1 + (1+r_{\rm onr} \Delta t) (C_0 - \delta S_0) - C_1,
\end{align}
where $S_1$ and $C_1$ denote tomorrow's prices of the underlying and the option, respectively, $r_{\rm onr}$ is the over-night rate at which the operator can borrow / lend money, and $\Delta t = 1/253$.   The operator' goal is to choose $\delta$ in such a way that the variance of tomorrow's wealth, ${\rm Var}[V_1^\delta]$ is minimised.

To make headway, since $\Delta t$ is small, we are allowed to approximate the variance by the expected squared mean. Indeed, if the expected return on the risky asset happens to be equal to the risk-free return then the expected value $\E[V_1^\delta]$ does not depend on $\delta$ at all.  Then the operator's objective is to minimise the \textit{mean squared hedging error} (MSHE) 
\begin{align} \label{eq:190407.1}
	\E \left[ \left(V_1^\delta\right)^2\right] = \E \left[ \left(\delta S_1 + (1+r_{\rm onr}  \Delta t) (C_0 - \delta S_0) - C_1\right)^2\right].
\end{align}
Let us assume for the moment that the option is a European call. Then a standard and simple choice is using  the  \textit{practitioner's Black-Scholes Delta} (BS Delta)
\begin{equation}
		 \label{eq:h_S}
	\delta_{\rm BS} = \mathbf{N}(d_1),
\end{equation}
	where $\mathbf{N}$ denotes the cumulative normal distribution function and 
	\begin{equation}  \label{eq:h_S'}
		d_1 = \frac{1}{\sigma_{\rm impl} \sqrt{\tau}} \left[ \ln\left(\frac{S_0}{K}\right) + \left(r + \frac{1}{2}\sigma_{\rm impl}^2\right)\tau  \right]. 		
\end{equation}
		Here,  $\tau$ is the time-to-maturity in year fraction, $\sigma_{\rm impl}$  the annualised implied volatility of the option, $K$  the strike price, and $r$ the risk-free interest rate corresponding to the option's maturity. The operator would choose $\delta = \delta_{\rm BS}$;  if the option was a put then she would choose $\delta = \delta_{\rm BS} - 1$ in line with put-call parity.  Since the interest rate $r$ is negligible, we assume for the moment that it is zero. Then 
the BS Delta can be written as a function of two variables, namely the moneyness $M = {S_0}/{K}$ and the square root of total implied variance $\sigma_{\rm impl} \sqrt{\tau}$. Thus, we get the functional representation
\[
	\delta_{\rm BS} = f_{\rm BS} \left(M, \sigma_{\rm impl} \sqrt{\tau}\right).
\]

It is now reasonable to study other functionals.
We shall replace $f_{\rm BS}$ by an ANN $f_{\rm NN}$ with the two input features $M$ and $\sigma_{\rm impl} \sqrt{\tau}$, trained to minimise the expression in \eqref{eq:190407.1}.  That corresponds to a nonparametric estimation of the optimal hedging ratio that minimises the variance of the hedging error. We will provide more details on the implementation in Section~\ref{sec:network}. The motivation to study ANNs arises from the large amount of historical data available, the universal approximation ability of ANNs, and  the sometimes unrealistic assumptions underlying parametric models.

To benchmark the hedging performance of the ANN, we introduce linear regression models that lead to hedging ratios that are linear in several option sensitivities. They are motivated by the \textit{leverage effect}, 
credited to  \cite{black1976studies}. The leverage effect
 describes the negative correlation between an underlying's price and its volatility. To illustrate how this matters, consider  a call and assume it is hedged with the BS Delta $\delta_{\rm BS} > 0$. If now the underlying's price goes up so do the call price and  the hedging position. Due to the leverage effect, the underlying (implied) volatility tends to go down simultaneously, thus having a negative effect on the option price. Indeed, everything else equal, both call and put prices go up as (implied) volatility increases -- their  `Vega' is positive. The  BS Delta $\delta_{\rm BS}$ does not take into consideration this additional effect. As we only allow hedging with the underlying the obvious change is to hedge only partially, i.e., use the hedging ratio $\delta_{\rm LR} = a \delta_{\rm BS}$, where $a$ is estimated (in a training set).  Here, ${\rm LR}$ stands for linear regression. For the moment it suffices to note that  these arguments let us expect $a >1$ for puts and $a<1$ for calls. (It turns out that hedging with $a \delta_{\rm BS}$, where $a = 0.9$ for calls  and $a=1.1$ for puts works extremely well on real-world datasets; see Subsection~\ref{SS:5.5}.) We shall discuss such simple modifications of the BS Delta in Section~\ref{S:benchmarks}, all based on statistical hedging models involving various option sensitivities.

The performance of the ANN and the benchmarks is tested on daily end-of-day mid-prices obtained from \textit{OptionMetrics} and tick data provided by \textit{Deutsche B\"orse}. These data are described in more detail in Section~\ref{S:data}. We also vary the length $\Delta t$ of the hedging period from 1 hour to 2 days.  All in all, the ANN performs  well in terms of MSHE relative to the BS Delta, even when the latter is being used with contract-specific implied volatility.  However, using the linear regression hedging ratios $\delta_{\rm LR}$ performs roughly as well or at times better than $\delta_{\rm NN}$. They lead to roughly 15\%-20\% reduction in the MSHE. For a summary of the results, see Section~\ref{S:Results}.  In addition, Online Appendix~\ref{A:simulation} contains an extensive simulation experiment using data generated from the standard Black-Scholes model and  from Heston's stochastic volatility model.

An interpretation of these observations is that the option sensitivities already encapsulate all relevant nonlinearities in the data necessary for the hedging task.   
 Hence, the ANN seems to be able to learn the leverage effect, but cannot improve on a simple linear regression involving the relevant option sensitivities. 
What have we learned? Initially we were satisfied about the outperformance of the ANN relative to the BS Delta on real-world datasets. When investigating what the ANN is learning, the linear regression models appeared as natural competitors. These statistical models are extremely simple -- for the  easiest such model one only replaces the BS Delta by a multiple of it. Nevertheless, as far as we know, these models have not been used in the literature to benchmark more complicated models.

We proceed as follows.  Section~\ref{S:data} describes the datasets and the experimental setup. Section~\ref{sec:network} introduces the HedgeNet architecture and implementation. This section also discusses the advantage of outputting directly the hedging ratio instead of option prices and then using a sensitivity as hedging ratio. Section~\ref{S:benchmarks} describes how the leverage effect motivates various benchmark models to be compared with ANNs. Section~\ref{S:Results} presents the experimental results.  Section~\ref{S:leakage} discusses potential information leakage introduced by the data cleaning procedure. 
\Cref{sec:conclusion} summarises the main findings. Several online appendices provide further details on the various sections.

\section{Datasets and setup of experiments} \label{S:data}
This section presents the data used.   Subsections~\ref{SS:SP500data} and \ref{SS:EuroStoxx}  describe the two real-world datasets containing options on the S\&P~500 and Euro Stoxx~50. Subsection~\ref{SS:preparation} discusses the experimental setup. 
Subsection~\ref{SS:MSHE} concludes the section by providing some economic implications of reducing the MSHE.
Online Appendix~\ref{A:sizes} contains additional details on these datasets. Online Appendix~\ref{A:simulation} discusses simulated datasets for an additional study.

\subsection{S\&P~500 end-of-day midprices} \label{SS:SP500data}
\begin{figure}[ht]
 	\centering
 	\includegraphics[width=1\linewidth]{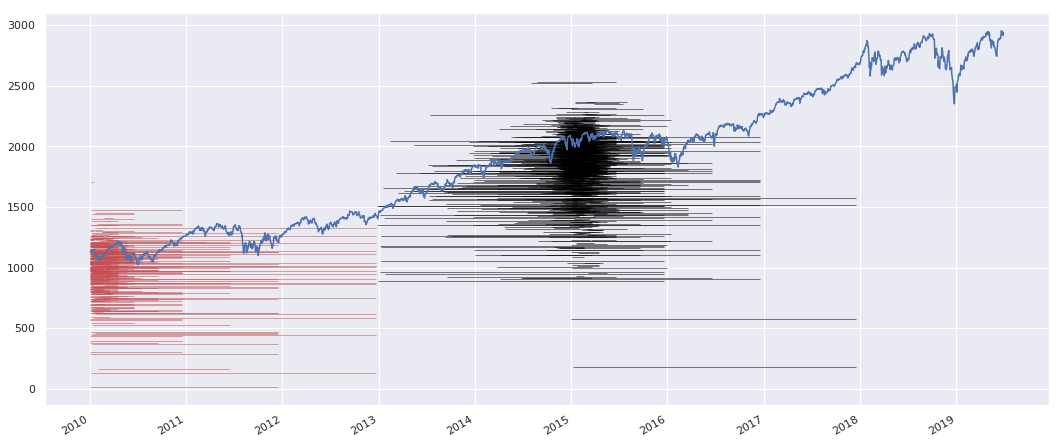}
 	\caption{
	A sample of the obtained put options along with the underlying's (S\&P~500) price process in blue. Only options that have a trading volume of more than 1000 on some trading day are included. Each red (black) line segment represents a put option that had price quotes within the first quarter of  2010 (2015).  The corresponding strike is indicated as  the value on the $y$--axis. Small random vertical shifts are added  to increase the visibility of the options.}
 	\label{fig:realoptions}
 \end{figure}

We obtained daily closing bid and ask prices on calls and puts written on the S\&P~500 
between January 2010 and June 2019  from 
 \textit{OptionMetrics} (see \url{https://optionmetrics.com}).  We interpret the midprice as the true market price.
 \Cref{fig:realoptions} displays a sample of the obtained options, namely those puts with price quotes in the first three months of 2010 or 2015.   Sensitivities are provided for the majority of options and are filled in for missing values. The results presented below are robust to whether we use computed sensitivities for all options  or the sensitivities provided by  \textit{OptionMetrics} where available. The required interest rates are interpolated from the rates provided by   \textit{OptionMetrics}. For maturities less than one week (in which case  \textit{OptionMetrics} does not provide the corresponding rates),  we use  the \textit{Overnight Libor Rates} from \textit{Bloomberg}.
 
We organised  the data in a table so that each row corresponds to exactly one observation, i.e.,  one option at one trading day (along with tomorrow's price for training).  We  remove certain samples; e.g., those samples with negative time-value, time-to-maturity less than 1 day, or zero trading volume. We present the full cleaning process in Online Appendix~\ref{A:cleaning}.

\subsection{Euro Stoxx~50 tick data} \label{SS:EuroStoxx}
We are grateful to \textit{Deutsche B\"orse}, who provided us with tick data of Euro Stoxx~50 index options and futures between January 2016 and July 2018. We refer to \url{https://datashop.deutsche-boerse.com/samples-dbag/File_Description_Eurex_Tick.pdf} for a description of this dataset.

We now briefly outline how we process these data. If several trades are executed at exactly the same time stamp we aggregate these orders and consider the volume-weighted average price.  We match each option transaction with the most recent tick price of the future with the shortest maturity (again, volume-weighted if several trades happen simultaneously). These futures, which are the most liquid ones, shall be used  to hedge the option position.
The computation of the option sensitivities requires a risk-free rate. We use interpolated Euro LIBOR rates from \textit{Thomson Reuters' DataStream}. 

To train the statistical models and to measure the hedging performance we require the option price after $\Delta t$ (1 hour, 1 day, 2 days, etc.). There might not be a trade exactly after this time period.  Hence we allow a \emph{matching tolerance window} of  6 minutes, equivalent to 0.1 hours.  Hence, for example, if $\Delta t$ is a business day and we have a trade on Monday, say at 2.12pm, then we match it with the first price observation of this option on Tuesday after 2.12pm. If there is no transaction before 2:18 pm, this sample gets discarded. We refer to Section~\ref{S:leakage} for a discussion of potential information leakage introduced in this step.

Finally, we perform a similar cleaning process as for the S\&P~500 dataset. The details are laid out again in Online Appendix~\ref{A:cleaning}.

\subsection{Data preparation and experimental setup} \label{SS:preparation}

As discussed in the introduction, our goal is to determine the hedging ratio $\delta$ as a function of observable quantities to minimise the variance over one period of the hedged portfolio
\begin{align} \label{eq:191020.1'}
V_1^\delta =  \delta S_1 + (1+r_{\rm onr} \Delta t) (C_0 - \delta S_0) - C_1.
\end{align}
Here $S_0$ and $S_1$ denote the prices of the hedging instrument at the beginning and end of the period and $C_0$ and $C_1$ denote the prices of the call or put.    We  study how well an ANN performs in this task on end-of-day midprices (see Subsection~\ref{SS:SP500data}) and on tick data (see Subsection~\ref{SS:EuroStoxx}).  We benchmark these results with linear regression models for the hedging ratio $\delta$.  A corresponding simulation study is discussed in Online Appendix~\ref{A:simulation}.

Each of the datasets is split up  into in-sample and out-of-sample (`test') data. Both the ANN and the benchmark models are trained to (estimated by) the in-sample dataset only.  The variance of the hedged portfolio is approximated by the MSHE. The performance of each of the methods is measured on the out-of-sample dataset as follows: 
\begin{align} \label{eq:200103}
\rm{Var} (V_1^\delta) \approx \rm{MSHE} = \frac{1}{N_{\rm test}} \sum_{t, j}^{N_{\rm test}}   \left(100 \frac{V_{t+1, j}^\delta}{S_{t}}\right)^2,
\end{align}
where $\delta$ is either modelled by an ANN or by a linear regression.
Both the indexing and the normalisation by $S_t/100$ need  explanation. 

First of all, the indexing has changed from \eqref{eq:191020.1'} to \eqref{eq:200103}. Indeed, each traded option  yields a series of samples, one for each trading period.  Moreover, several options corresponding to different strikes (indexed by $j$) are being priced in any  specific period (e.g., a day). To emphasise this point, the samples are double indexed in \eqref{eq:200103}.
Next, \eqref{eq:200103} normalises the value of the hedging portfolio by dividing it by $S_t/100$.  This normalisation `removes the units' and allows to compare errors across the different datasets, and arguably more importantly, across time. 
Equivalently, at any point of time $t$, instead of replicating a full option we replicate the fraction $100/S_t$ of this option.  

One could have considered a different normalisation. For example, in \eqref{eq:200103}, one could have divided by the time-$t$-option price $C_t$ instead of $S_t$. This would induce a different weighting of the samples. However, a fixed Dollar position in a far out-of-the money option is riskier than in an at-the-money option. Indeed, a move in the underlying tends to have a larger effect on the far out-of-the money position. 
Hence from a risk perspective, the alternative normalisation would put too much weight on  far out-of-the money options.  For this reason we choose the normalisation of \eqref{eq:200103}.

We now provide more details on how we prepare each dataset. First we store each dataset in a dataframe as in Table~\ref{tab:dataset_simulation}. We then remove all in-the-money samples. That is, if at one specific date an option was in the money, we discard this specific date for the corresponding option.

\begin{table}[ht] 
	\begin{tabularx}{\textwidth}{c c *{10}{Y}} 
		\toprule
		Index
		& Date
		& \multicolumn{4}{c}{Features}  
		& \multicolumn{5}{c}{Additional information} & Target\\
		\cmidrule(lr){3-6} \cmidrule(l){7-11}
		& & $\sigma_{\rm impl} \sqrt{\tau}$ & $M$ & $\delta_{\rm BS}$ & ${\mathcal V}_{\rm BS}$ & $S_0$ & $S_1$ & $C_0$ & $r_{\rm onr}$ & CP flag& $C_1$ \\
		\midrule
		0 & 2018/07/02 & $0.047$ & $1.003$ & $0.531$ & $9.357$ & $100$ & $98.223$ & $2.002$ & $0.01$ & $0$ & $1.130$ \\
		$\vdots$  \\
		\bottomrule
	\end{tabularx}
	\caption{This table presents a (simplified) preview of one of the four processed datasets.   The `Features' columns are used as inputs for the ANN and the linear regressions.  The labels $\sigma_{\rm impl} \sqrt{\tau}$ and $M$ denote the square root of total implied variance and moneyness of the option. The labels $\delta_{\rm BS}$ and ${\mathcal V}_{\rm BS}$  are the BS Delta and Vega.  The CP flag indicates whether the corresponding option is a call or a put. Prices and sensitivities are all normalised.}
	\label{tab:dataset_simulation}
\end{table}

We break up the S\&P~500 dataset  in 14 overlapping time windows of length 3 years in order to understand whether the comparisons between the ANNs and the linear regressions are consistent across time.  
In each time window, the first 900 days form the in-sample set, while the last 180 days are used for the out-of-sample set, yielding a ratio 5:1.  For the training of the ANN, the 900  days are furthermore split into 720  days of training and and 180  days of validation yielding a ratio 4:1:1. We roll the time windows forward by 180 days, so that sample appears maximally once in the aggregated out-of-sample set.
The Euro Stoxx~50 dataset is much shorter, and we do not break it up in different time windows. 
This leads to 750 (600+150) days in the in-sample set and 150 days in the out-of-sample set, yielding again a  ratio 4:1:1. 

In practice, one would expect to retrain each statistical model weekly or daily instead of every 180 days as done in the S\&P dataset. For computational limitations we are not able to do so. (Currently, training and running one ANN configuration for the 14 S\&P time windows  takes about 10 hours on a GTX 1060 6GB GPU cluster.) We treat the statistical benchmark models below in the same way, also only retraining them every 180 days.

\subsection{Digression: economic interpretation of the mean squared hedging error} \label{SS:MSHE}
We now briefly comment on the economic gains when using hedging strategies that lead to reduced MSHEs.
We have in mind a financial entity (or `operator')  acting as a market maker; i.e., taking on (short) positions in options as `inventory' to satisfy some market demand. 
 This operator sells a cross section of delta-hedged puts or calls. In the classical one-period framework of \cite{stoll1978supply} (see also Chapter~2.2 in \cite{ohara1997market}), the operator charges a premium (e.g., through a bid-ask spread) to take on the additional inventory (i.e., the short position of delta-hedged options).  Reducing the MSHE allows the operator to charge a lower premium as we outline next.

Formally, we equip the operator with quadratic utility $x \mapsto x- \gamma x^2/2$, where $\gamma>0$ denotes her coefficient of risk aversion.  We suppose that the delta-hedged short-position is uncorrelated with the operator's optimal wealth.  Furthermore, we assume that the expected return of a delta-hedged option position does not depend on the hedging strategy (e.g., if the expected return of the risky asset equals the risk-free return) and set it to zero for simplicity.  Under Bertrand competition of liquidity providers with the same risk aversion $\gamma$, the operator charges $\gamma/2$ times the MSHE as a premium.  Hence, if the MSHE can be reduced by a certain percentage, the premium reduces by the same percentage times $\gamma/2$. For example, if the MSHE error is reduced by 15\% and $\gamma = 2$ then the premium decreases by 15\%.

A similar argument applies if the financial entity was  on the `buy-side,'  taking on short positions in options to collect the volatility risk premium, and  interested in maximising the Sharpe ratio of her position.  This entity would then try to hedge the exposure to the price movements in the underlying by trading it. 
If the expected return of a delta-hedged option position does not depend on the hedging strategy and the MSHE is reduced by 15\% then the new Sharpe ratio is $1/\sqrt{0.85}  \approx 1.085$ times the old one.

\section{HedgeNet} \label{sec:network}
There exists a long line of research on the use of ANNs in the context of option pricing and hedging.   \cite{RW.2019.LR} provide an overview of this literature. Here we only give a few pointers to papers that we found especially insightful. Early on, \cite{hutchinson1994nonparametric} suggest ANNs as nonparametric alternative for the pricing of options. They show that already quite small ANNs with only a few nodes   perform well for the pricing task. 
	 \cite{garcia2000pricing} are among the first to introduce financial domain knowledge (a so called `homogeneity hint') in the design of ANNs. This type of regularisation improves the pricing performance of ANNs further.	
	 \cite{carverhill2003alternative}  propose an ANN that directly outputs hedging strategies, instead of  first outputting option prices and then deriving  hedging strategies as sensitivities. 
	 \cite{dugas2009incorporating} suggest an ANN architecture that guarantees that the outputted prices satisfy a set of no-arbitrage conditions.  	
	 \cite{buehler2019deep} bring several innovations forward. In order to train their ANN, additional artificial data are drawn from an appropriately fitted econometric model. Their framework for hedging options includes the presence of transaction costs and other market frictions, allowing general convex risk measures as loss functions.  All these references discussed here consider the pricing / hedging task over the lifespan of an option.
	 
We now introduce the ANN used in this study. As discussed in the introduction, we focus on the one-period setup, and benchmark the hedging performance of the ANN with appropriate linear regressions based on the options' sensitivities, as described in the next section. 	 
The ANN  maps an option's relevant features (e.g, moneyness and square root of total implied variance) to a hedging ratio $\delta_{\rm NN}$.  In Subsection~\ref{SS:architecture}  we provide details about the architecture, implementation, and training of such an ANN. 	 
Subsection~\ref{SS:Why} provides some additional motivation why the ANN is designed to output directly the hedging ratio instead of the option price.

\subsection{Architecture of HedgeNet, its implementation and training} \label{SS:architecture}
An ANN is a composition of simple elements called neurons,  which maps input features to outputs. Such an ANN then forms a directed, weighted graph.

\begin{figure}[ht]
	\centering

\tikzstyle{input} = [rectangle, rounded corners, minimum width=3cm, minimum height=1cm,text centered, draw=black, fill=red!30]

\tikzstyle{output} = [rectangle, rounded corners, minimum width=3cm, minimum height=1cm,text centered, draw=black, fill=red!30]

\tikzstyle{process} = [rectangle, minimum width=3cm, minimum height=1cm, text centered, draw=black, fill=orange!30]

\tikzstyle{transform} = [rectangle, minimum width=3cm, minimum height=3cm, text centered, text width=3cm, draw=black, fill=black!30]

\tikzstyle{arrow} = [thick,->,>=stealth]
\begin{tikzpicture}[scale=0.5,node distance=2cm]

\node (input1) [input] {Features};
\node (input2) [input, below of=input1] {Additional inputs};
\node (fcnn) [process, right of=input1, xshift=2.2cm] {FCNN};
\node (transform) [transform, right of=fcnn, yshift=-1cm, xshift=2.2cm] {Nontrainable transformations};
\node (output) [output, right of=transform, xshift=2.2cm] {$\widehat{C}_1$};

\draw[arrow] (input1) -- (fcnn);
\draw[arrow] (fcnn) -- (fcnn-|transform.west);
\draw[arrow] (input2) -- (input2-|transform.west);
\draw[arrow] (transform) -- (output);

\end{tikzpicture}
\caption{A schematic graph of HedgeNet. The features are transformed into a hedging position by a fully-connected feed-forward neural network (FCNN). The additional input is used to compute the value $\widehat C_1$ of the hedging position.  }
\label{fig:Hedgenet}
\end{figure}
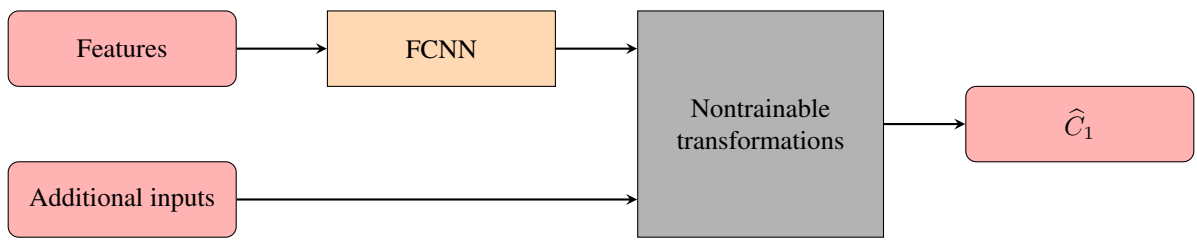

As we shall discuss below in Subsection~\ref{SS:Why} it is not satisfactory to compute or estimate option prices and then use their  sensitivities as hedging ratios.  It is better to obtain the hedging ratio, our quantity of interest, directly. Hence, we desire that the ANN returns a hedging ratio and not a price. However, when training such an ANN what should it be trained to? Optimal hedging ratios are not provided in the data. For this reason, we design an ANN, named HedgeNet, to have two parts, as illustrated in \Cref{fig:Hedgenet}.

The first part, a multilayer fully-connected feed-forward neural network (FCNN), transforms features into a hedging position, which is then turned by the second part into the  replication value $\widehat C_1 = V_1 + C_1$. This output of HedgeNet can then be trained to the observed option prices $C_1$ at the end of each period by minimising the sum of squared differences.

The FCNN has two hidden layers with 30 nodes each, connected by ReLU activation. (The benefits of using ReLU activation are addressed in \cite{glorot2011deep} and Section~3.1 of \cite{krizhevsky2012imagenet}.)
The output of the FCNN  is provided by a linear node (with truncation at zero and one) and corresponds to the the hedging ratio $\delta_{\rm NN}$. We  tried different architectures, for example 100 nodes in each hidden layer, or three (instead of two) hidden layers with 30 nodes each. Motivated by the representation of the BS Delta in \eqref{eq:h_S}, we also tried the cumulative distribution function $\mathbf{N}$ of a standard normally distributed random variable as output function instead of the linear output function. None of these modifications changed the overall conclusions below.  We also tried a modification, where we interpret the output not as the hedging ratio but as the `bias' term $\delta - \delta_{\rm BS}$, which corrects the BS Delta.  Such change did not help the performance of the ANN either.

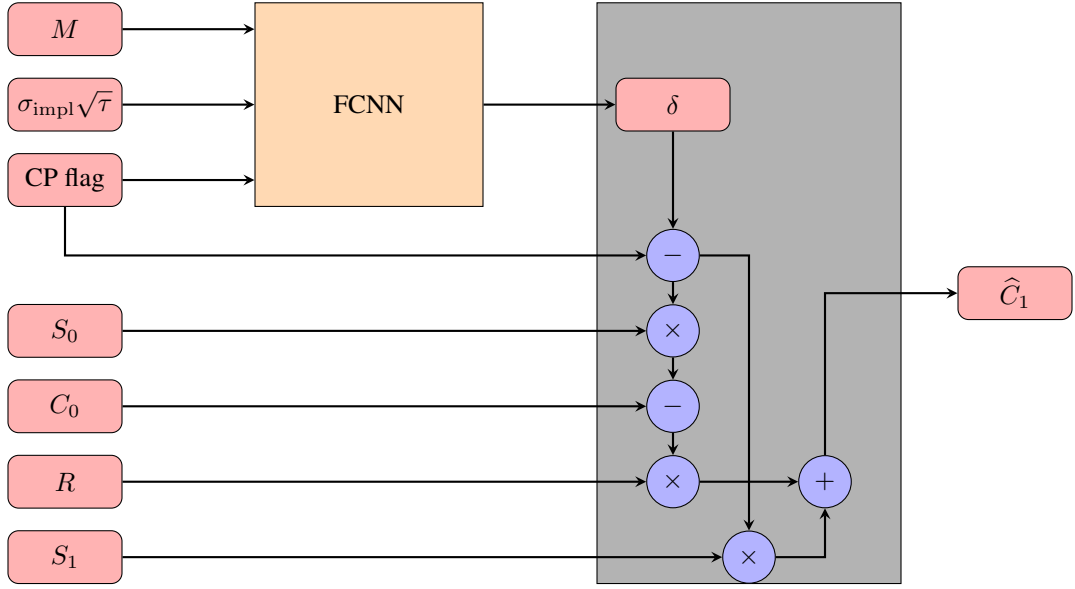
\begin{figure}[ht]
	\centering
	\tikzstyle{var} = [rectangle, rounded corners, minimum width=1.5cm, minimum height=0.7cm,text centered, draw=black, fill=red!30]
	
	\tikzstyle{fcnn} = [rectangle, minimum width=3cm, minimum height=2.7cm, text centered, draw=black, fill=orange!30]
	
	\tikzstyle{transform} = [rectangle, minimum width=4cm, minimum height=7.7cm, text centered, text width=3cm, draw=black, fill=black!30]
	
	\tikzstyle{op} = [circle, radius=0.35cm, text centered, draw=black, fill=blue!30]
	
	\tikzstyle{arrow} = [thick,->,>=stealth]
	
	\begin{tikzpicture}[node distance=1cm]
	
	\node (moneyness) [var] {$M$};
	\node (sigtau) [var, below of=moneyness] {$\sigma_{\rm impl}\sqrt{\tau}$};
	\node (cp) [var, below of=sigtau] {CP flag};
	
	\node (s0) [var, below of=cp, yshift=-1cm] {$S_0$};
	\node (c0) [var, below of=s0] {$C_0$};
	\node (R) [var, below of=c0] {$R$};
	\node (s1) [var, below of=R] {$S_1$};
	
	\node (fcnn) [fcnn, right of=sigtau, xshift=3cm] {FCNN};
	
	\node (transform) [transform, right of=fcnn, xshift=4cm, yshift=-2.5cm] {};
	\node (delta) [var, right of=fcnn, xshift=3cm] {$\delta$};
	\node (minus-cp) [op, below of=delta, yshift=-1cm] {$-$};
	\node (times-s0) [op, below of=minus-cp] {$\times$};
	\node (c0-minus) [op, below of=times-s0] {$-$};
	\node (times-r) [op, below of=c0-minus] {$\times$};
	\node (plus-delta-s1) [op, right of=times-r, xshift=1cm] {$+$};
	\node (times-s1) [op, below of=times-r, xshift=1cm] {$\times$};
	
	\node (c1) [var, right of=transform, xshift=2.5cm] {$\widehat{C}_1$};

	\draw[arrow] (moneyness) -- (moneyness-|fcnn.west);
	\draw[arrow] (sigtau) -- (sigtau-|fcnn.west);
	\draw[arrow] (cp) -- (cp-|fcnn.west);
	
	\draw[arrow] (fcnn) -- (delta);
	\draw[arrow] (delta) -- (minus-cp);
	\draw[arrow] (cp) |- (minus-cp);
	\draw[arrow] (minus-cp) -- (times-s0);
	\draw[arrow] (s0) -- (times-s0);
	\draw[arrow] (times-s0) -- (c0-minus);
	\draw[arrow] (c0) -- (c0-minus);
	\draw[arrow] (c0-minus) -- (times-r);
	\draw[arrow] (R) -- (times-r);
	\draw[arrow] (s1) -- (times-s1);
	\draw[arrow] (minus-cp) -| (times-s1);
	\draw[arrow] (times-r) -- (plus-delta-s1);
	\draw[arrow] (times-s1) -| (plus-delta-s1);
	\draw[arrow] (plus-delta-s1) |- (c1);
	
	\end{tikzpicture}
	\caption{A  detailed schematic presentation of HedgeNet.  Recall that $M = S_0/K$ and $\sigma_{\rm impl} \sqrt{\tau}$ are moneyness and square root of total implied variance. `CP flag' is a Boolean flag for the option type; it equals 1 for puts and 0 for calls. Next, $S_0$ and $S_1$ are the underlying's prices at the beginning and end of the hedging period, $C_0$ denotes the option price  at the beginning of the period,  and $\widehat C_1$ denotes the replication value. Finally, $R=1+r_{\rm onr} \Delta t$ is the risk-free overnight return.}
	\label{fig:hedgenet_concrete}
\end{figure}

As illustrated in Figure~\ref{fig:hedgenet_concrete}, the non-trainable transformation module turns the hedging ratio  $\delta_{\rm NN}$ into the replication value $\widehat C_1$ by following \eqref{eq:191020.1}.
As the data includes both puts and calls, this module also requires an option type flag, which is set to $1$ in the case of a put and to $0$ in the case of a call.  If the sample is a put, the module replaces $\delta_{\rm NN}$  by $\delta_{\rm NN}-1$  in line with put-call parity.   The non-trainable transformation module consists of a series of affine transformations, and hence does not affect the universal approximation property, discussed for example in \cite{yarotsky2017error}.

All numerical experiments  are run on a standard desktop with GPU accelerated computation (specification: GTX 1060 6GB GPU). 
We use \textit{Python} as  programming language. The ANN is implemented with the  deep learning framework \textit{Tensorflow} along with \textit{Keras}.  
The inputs to the trainable part of HedgeNet are standardised.  The weights of the ANN are initialised via the `Xavier' initialiser (\cite{glorot2010understanding}) and the 
 `Adam' optimiser (\cite{kingma2014adam}) is applied for training the ANN. Online Appendix~\ref{A:reg-para} contains details on the choice of additional hyperparameters.

For each dataset we consider three different feature sets for the trainable part of HedgeNet:
\begin{itemize}
	\item $\text{ANN}(M;\, \sigma_{\rm impl} \sqrt{\tau})$: The first one is already indicated in Figure~\ref{fig:hedgenet_concrete}. It uses moneyness $M$, square root of total implied variance $\sigma_{\rm impl} \sqrt{\tau}$, and a flag to indicate whether the option is a call or a put.    
It is worth pointing out that using  moneyness instead of  the underlying's price and the strike price separately offers a better generalisation performance. The most important reason for its better performance is that  moneyness resembles more a stationary feature  compared to the underlying's price and strike price separately.  Indeed, options are created and traded only for a certain range of moneyness values.  \cite{ghysels1997nonparametric}, \cite{garcia2000pricing}, and \cite{RW.2019.LR} provide more comments on the advantage of using moneyness.   
The choice of square root of total implied variance is motivated by the fact that volatility squares with the square root of time; see also the expression for $\delta_{\rm BS}$ in \eqref{eq:h_S}\&\eqref{eq:h_S'}.

\item $\text{ANN}(\Delta_{\rm BS};\, \mathcal{V}_{\rm BS};\, \tau)$: Motivated by the leverage effect discussed in Section~\ref{S:benchmarks} below, we also consider a second set of features consisting of $\delta_{\rm BS}$, $\mathcal V_{\rm BS}$,  $1/{\sqrt{\tau}}$, and the put-call flag. Here $\mathcal V_{\rm BS}$ denotes \textit{Vega}, the sensitivity of the option price with respect to the implied volatility.

\item  ANN($\Delta_{\rm BS};\, \mathcal{V}_{\rm BS};\, {\rm Va}_{\rm BS};\, \tau$): Since we shall use Vanna, the sensitivity of Delta with respect to volatility,  as a feature for linear regression benchmarks in Section~\ref{S:benchmarks}, we also consider using a third feature set consisting of the three sensitivities, $1/{\sqrt{\tau}}$, and the put-call flag.
\end{itemize}

\subsection{Digression: Why outputting the hedging ratio instead of computing price sensitivities?} \label{SS:Why}

Most  ANNs constructed in the literature for the risk management of options first learn the pricing function. Then in a second step  hedging strategy is computed as the sensitivity of the option price with respect to the underlying's price; see  \cite{RW.2019.LR} for an overview of the literature. In contrast, HedgeNet
allows to predict the hedging position directly. In this way, the hedging strategy is no longer interpreted as a sensitivity.

From a risk-management point of view the hedging ratio is the main quantity of interest.  It is recommended, see for example \cite{Bengio:1997} or \cite{Claeskens:2003},  to estimate  relevant quantities directly.  This is in line with the important observation made in  \cite{Lyons_1995} that different models might yield similar option prices but completely different hedging strategies. 
Obtaining directly the hedging ratio  also avoids the otherwise necessary step to differentiate, possibly numerically, the trained option prices. 

There are further important advantages of outputting directly the hedging ratio. Computing sensitivities usually does not take into consideration that other model parameters also might change, in line with the underlying. Hence, such sensitivities tend to be not optimal for reducing the MSHE. Theoretical results supporting this observation are ample; see for example \cite{denkl2013performance}.  This discussion is continued in Subsection~\ref{SS:Delta-vega} below. Moreover, as \cite{buehler2019deep} show, training to hedging ratios allows to incorporate market frictions conveniently. 

At this point, let us also mention a different approach to use ANNs in the context of option pricing, namely as computational tools to replace expensive PDE solvers or Monte-Carlo simulations.  Indeed the risk management of `sell-side institutions' is  subject to regulatory purposes. In particular, their options' hedging is supposed to be derived from specific parametric models. ANNs are used to estimate (`calibrate') these model parameters. 
For references using this approach, see  \cite{RW.2019.LR}.
Here, however, we do not intend to study the question how well models can be calibrated by the use of ANNs. Instead, we show the limitations and benefits of ANNs for estimating the optimal hedging ratio when not being restricted by a specific parametric model.

\section{Linear regression models as benchmarks} \label{S:benchmarks}
We now discuss how we benchmark the hedging performance of the ANN.  Although not very reasonable, one benchmark could be not hedging at all, i.e., $\delta = 0$. In this case the variance of the hedging error  is just the variance of the change in the option price.  More reasonable is to use the BS Delta, obtained from the Black-Scholes formula, as discussed in Subsection~\ref{SS:benchmark BS}.
Subsections~\ref{SS:Delta-vega} and \ref{SS:other benchmarks} introduce some further simple statistical hedging models. 

\subsection{Black-Scholes benchmark} \label{SS:benchmark BS}
Hedging via the BS Delta is a standard benchmark. That is, for each option and for each date the corresponding implied volatility is used to obtain the hedge in \eqref{eq:h_S}, namely the partial derivative of the Black-Scholes option price with respect to the price of the underlying. Black-Scholes performs best if implied volatility is plugged in.  In the literature, other volatilities, such as historical volatility estimates  or  GARCH predicted volatilities have been used. We refer to \cite{RW.2019.LR} for an overview.

Since here we hedge only discretely, using the BS Delta  leads to an error even if the data are simulated from the Black-Scholes model.
The performance of discrete-time hedging  has been extensively studied; some pointers to the literature include  \cite{boyle1980discretely}, \cite{bertsimas2000time}, and \cite{tankov2009asymptotic}, who provide an asymptotic analysis of hedging errors.

\subsection{Delta hedging other sensitivities} \label{SS:Delta-vega}
The leverage effect, first discussed in \cite{black1976studies}, describes the negative correlation of observed returns and their volatilities in equity markets. This effect has been confirmed in many follow-up studies which also consider implied volatilities. For example,  \cite{cont2002dynamics} claim that the leverage effect is due to a shift in the overall level of the implied volatility surface and not due to relative movements, that is, changes in the shape of the implied volatility surface.  The non-zero correlation of returns and the implied option volatilities indicates that the BS Delta  can usually be outperformed by some relatively simple adjustments.
In this spirit, \cite{vahamaa2004delta} and \cite{crepey2004delta} use the observed smile in option implied volatilities to improve on the hedging performance of the BS Delta. These ideas are developed further in several papers; see for example, \cite{alexander2012regime}.

The central idea is to note that a first-order Taylor series expansion of option prices yields 
\[
	\d C \approx \delta_{\rm BS}\, \d S + \mathcal{V}_{\rm BS} \, \d \sigma_{\rm impl}  = \delta_{\rm BS}\, \d S + \mathcal{V}_{\rm BS} \,\frac{\d \sigma_{\rm impl}}{\d S} \,\d S + \mathcal{V}_{\rm BS} \,\d S^\bot,
\]
where $S^\bot$ is orthogonal to $S$. In words, the change in the option price is approximately the BS Delta times the change in the underlying's price plus Vega times the change in the implied volatility.  The second term can be written in terms of changes in the underlying's price and changes in the implied volatility that are uncorrelated with the changes in the underlying's price.
These observations lead us to consider a statistical model of the form:
\begin{align*}
		\delta =  a\, \delta_{\rm BS} + b\, \mathcal{V}_{\rm BS}.
\end{align*}
This statistical model replaces the BS Delta by a multiple $a$ of it plus a multiple $b$ of Vega $ \mathcal{V}_{\rm BS}$. Here, $a$ and $b$ are estimated in the in-sample set, separately for puts and calls. More precisely, estimating $a$ and $b$ is equivalent to running a linear regression with two independent variables and no intercept on the in-sample set. 
Indeed, we minimise the expression in \eqref{eq:200103}, where each  summand  can be written as the square of 
\[
	a \left(\delta_{{\rm BS}, t, j} \, x_t\right) + b \left( \mathcal{V}_{{\rm BS},t,j} \, x_t\right)   -  y_{t,j},
\]
with $x_t = 100 (S_{t+1}/S_t  -  (1+r_{\rm onr}  \Delta t)  )$ and $y_{t, j} =  100 / S_t (C_{t+1,j} - (1+r_{\rm onr}  \Delta t) C_{t,j})$.

Next, a Taylor series expansion of the BS Delta yields 
\begin{align*}
	\d \delta  \approx  \Gamma_{\rm BS}\, \d S + {\rm Va}_{\rm BS}\, \d \sigma_{\rm impl}.
\end{align*}
Here, $ \Gamma_{\rm BS}$ denotes  \textit{Gamma}, namely the sensitivity of the BS Delta to changes in the underlying's price; ${\rm Va}_{\rm BS}$ denotes \textit{Vanna}, namely the sensitivity of the BS Delta to changes in the implied volatility. 

Combining these two expansions we obtain the linear regression model
\begin{align} \label{eq:200118}
	\delta_{\rm LR} = a \,\delta_{\rm BS} + b\, \mathcal{V}_{\rm BS} + c\, {\rm Va}_{\rm BS} + d  \,\Gamma_{\rm BS}.
\end{align}
Again, $a,b,c,d$ are estimated for puts and calls separately on each in-sample set. We also consider nested models; in this case, we force either $a$ to be one or one (or more) of the other coefficients to be zero and estimate the remaining coefficients.  The Vega and Gamma sensitivities are large for options when the strike is close to the underlying's current price. Thus, including these sensitivities allow the statistical model to make adjustments to the hedging ratio depending on whether an option is at-the-money or out-of-the money.  Using both two sensitivities helps, moreover, to make additional adjustments depending on the option's time-to-maturity. Finally, Vanna for an out-of-the money option is largest when the option is somehow out-of-the-money but not too much. This allows the model to make the corresponding additional adjustments. 
We have also experimented with an additional intercept term in \eqref{eq:200118}. Including it does not change the conclusions below; we hence only report the results without this additional term.

Furthermore, we include below the proposed hedging ratio of \cite{hull2017optimal}, given by 
\begin{align} \label{eq:HW}
	\delta_{\rm HW} = \delta_{\rm BS} + \frac{ \mathcal{V}_{\rm BS}}{\sqrt{\tau} S} (a + b  \delta_{\rm BS} + c\delta_{\rm BS}^2).
\end{align}
Here, $\tau$ is the time-to-maturity and $a,b,c$ are again estimated  for puts and calls separately on each in-sample set.
\cite{hull2017optimal} obtain this  model from a careful analysis of S\&P~500 options and observe its  excellent hedging performance on options written on the S\&P~500 and other indices.  We furthermore include a `Relaxed Hull-White' model, where the coefficient in front of $\delta_{\rm BS}$ is not restricted to one.

The models in  \eqref{eq:200118} and \eqref{eq:HW} should  be considered `statistical' in contrast to `model-driven' as the hedging ratio is derived purely from statistical considerations instead of being derived from stochastic models.  In the language of \cite{Carr.Wu.2019}, these  models are `local' and `decentralised,' as only  one period is considered instead of the option's whole time horizon, and as  each option contract is treated separately instead of finding an overall consistent valuation model.
To the best of our knowledge, the model in \eqref{eq:200118} has not been suggested in the literature before, despite its simplicity. (Relatedly, \cite{Bergomi2009} introduces the `skew stickiness ratio' to describe the idea that changes in the at-the-money implied volatility relative to the underlying's logarithmic return is proportional to the implied at-the-money volatility skew. The proportionality constant can then be estimated again by linear regression. In the context of credit risk, \cite{ContKan} also provide a careful study of regression-based hedging. While here the hedging ratio is regressed on option sensitivities, they regress  changes in the option price on changes in the underlying.)

\subsection{Possible other benchmarks} \label{SS:other benchmarks}
One could consider hedging ratios derived from parametric models such as stochastic volatility models.  \cite{bakshi1997empirical} observe that such models outperform the BS Delta in the case of hedging out-of-the money options, but not necessarily in-the-money options.   \cite{vahamaa2004delta} provides additional references that test the hedging performance of stochastic volatility models and concludes with the observation that ``such models do not necessarily provide better hedging performance.'' \cite{hull2017optimal} note that  the hedging ratio $\delta_{\rm HW}$ of \eqref{eq:HW} leads to a better performance than stochastic volatility models.

We initially also investigated the following two  (semi-)linear benchmarks:
	\begin{equation*}
	\overline{\delta}_1 = a M + b\sigma_{\rm impl}\sqrt{\tau} + c; \qquad \overline{\delta}_2 = \mathbf{N}\left(a M + b\sigma_{\rm impl}\sqrt{\tau} + c\right),
	\end{equation*}
	where $M$ denotes moneyness, $\sigma_{\rm impl} \sqrt{\tau}$  square root of total implied variance,  and $\mathbf{N}$  the cumulative normal distribution function.  Here, the parameters $a,b,c$ were estimated again in each in-sample set. 
	It turns out that  these two linear regressions perform far worse than the BS Delta $\delta_{\rm BS}$; hence we will not present results on these two benchmarks. The underperformance of these two linear regressions also shows that the performance of the ANN is not entirely due to the hand-crafted features.

\section{Results} \label{S:Results}

We now present the results on the performance of the various statistical hedging models  in terms of MSHE reduction. As a quick summary, the hedging ratios of the ANNs do not outperform the linear regression models.  On the S\&P~500 dataset, the Hull-White and Delta-Vega-Vanna regressions tend to perform the best, with Hull-White better on the one-day hedging period, and the Delta-Vega-Vanna regression better on the two-day period.  On the Euro Stoxx~50 dataset, the Delta-Vega-Gamma-Vanna regression tends to perform the best. However, the differences between these linear regressions with three or four coefficents are neither statistically nor economically significant, as we shall discuss.

Recall from Subsection~\ref{SS:preparation} that each data sample is normalised so that the underlying's price $S_0$ at time 0 is 100.  This allows to compare the absolute hedging errors across different datasets.  Recall also that we only consider out-of-the (and at-the)-money puts and  calls. In the next two subsections we discuss the results for the S\&P~500 and Euro Stoxx~50 datasets.  In Subsection~\ref{SS:5.5}, we conclude this section with some general observations and guidelines.

\subsection{S\&P~500 end-of-day midprices} \label{SS:SP500}
\Cref{tab:sp-overall} gives an overview of the MSHEs across different hedging periods.  
The first two rows give the MSHEs for the zero hedge and the BS Delta. The remaining rows give the relative improvement over the BS Delta, i.e., 
 \begin{equation} \label{eq:relative-improvement'}
\frac{\text{MSHE} (\delta_*) - \text{MSHE} (\delta_{\rm BS})}{\text{MSHE} (\delta_{\rm BS})},
 \end{equation}
 
 All  competing methods outperform the BS Delta.  Among them, the Delta-Vega-Vanna and (relaxed) Hull-White regressions perform the best, with Hull-White doing slightly better on the one-day hedging period while Delta-Vega-Vanna performing better on two-day hedging period. Indeed,  \cite{hull2017optimal}  study the same dataset to create the Hull-White regression, so it is surprising how close the other regressions get.   The major improvement in the regressions (apart from the Hull-White regression) comes from allowing the coefficient in front of Delta  to be  estimated, rather than equal to one. 
 The ANNs perform similarly to the regressions in case of the one-day period, but underperform for the two-day period.

\begin{table}[ht]
	\centering
	\begin{tabular}{c c|c c c |c c cl}
	& \multirow{2}{*}{} & \multicolumn{3}{c|}{1 day} & \multicolumn{3}{c}{2 days}\\
	
	& & Calls & Puts & Both & Calls & Puts & Both \\
	\cmidrule{2-9}

	& Zero hedge & 4.01 & 4.78 & 4.54 & 8.31 & 9.73 & 9.29\\
	& BS Delta  & 0.687 & 0.655 & 0.665 & 1.58 & 1.54 & 1.55\\
	\cmidrule{2-9}
	\multirow{3}{*}{\hspace{-1em}$\left.\begin{array}{l}
		\\
		\\
		\\
		\\
		\\
		\\
		\\
		\\
		\\
		\\
		\\
		\\
		\\
		\end{array}\text{Regressions}\right\lbrace$}
	& Delta-only & -21.3 & -14.8 & -16.9 & -16.3 & -12.8 & -13.9 & \\
	& Vega-only & -13.7 & -11.7 & -12.3 & -10.4 & -10.1 &-10.2\\
	& Gamma-only & -15.5 & -10.1 & -11.8 & -14.5 & -11.2 &-12.2\\
	& Vanna-only & -12.4 & -12.6 & -12.5 & -10.6 & -13.0 &-12.2\\
	& Delta-Gamma & -21.6 & -14.8 & -17.0 & -17.1 & -13.1 & -14.4 \\
	& Delta-Vega & -21.4 & -14.9 & -17.0 & -16.4 & -12.8 & -13.9 \\
	& Delta-Vanna & -22.6 & \textbf{-16.6} & \textbf{-18.5} &\textbf{-17.7} & \textbf{-15.4} & \textbf{-16.1} \\

	& Delta-Vega-Gamma & -21.5 & -14.8 & -17.0 & -16.8 & -13.5 & -14.5\\
	& Delta-Vega-Vanna & \textbf{-23.0} & \textbf{-16.6} & \textbf{-18.7} & \textbf{-18.1} & \textbf{-15.4} & \textbf{-16.2} \\
	& Delta-Gamma-Vanna & -22.6 & \textbf{-16.6} & \textbf{-18.5} & \textbf{-17.7} & \textbf{-15.2} & \textbf{-16.0}\\
	& Delta-Vega-Gamma-Vanna & \textbf{-22.9} & \textbf{-16.4} & \textbf{-18.5} & -17.4 & \textbf{-14.9} & \textbf{-15.7} \\ 
	& Hull-White  & \textbf{-23.1} & \textbf{-16.9} & \textbf{-18.9} & \textbf{-17.8} & \textbf{-14.5} & \textbf{-15.5}\\
	& Relaxed Hull-White & \textbf{-23.2} & \textbf{-16.9} & \textbf{-18.9} & \textbf{-18.3} & \textbf{-14.6} & \textbf{-15.8} \\
	\cmidrule{2-9}
		\multirow{2}{*}{\hspace{1em}$\left.\begin{array}{l}
			\\
			\\
			\\
			\end{array}\text{ANNs}\right\lbrace$}
	& $M;\, \sigma_{\rm impl} \sqrt{\tau}$ & -22.3 & -15.6 & -17.7 & -17.1 & -10.9 & -12.8\\
	& $\Delta_{\rm BS};\, \mathcal{V}_{\rm BS};\, \tau$ & \textbf{-23.4} & \textbf{-16.9} & \textbf{-18.9} & \bf-18.6 & -12.9 & -14.7
	 \\
	& $\Delta_{\rm BS};\, \mathcal{V}_{\rm BS};\, {\rm Va}_{\rm BS};\, \tau$
	& -21.9 & -14.4 & -16.8 & -12.5 & -12.9	 & -12.8
	\end{tabular}
	\caption{Performance of the linear regressions and ANNs on the S\&P~500 dataset.  The hedging periods $\Delta t$ are here either one day or two days.  The columns `Both' are the weighted average of the `Puts' and `Calls' columns. The row `Zero hedge' corresponds to the MSHE when $\delta=0$ is chosen; i.e., the mean squared changes in the option prices.  The values in the top two rows are multiplied by 100 to improve readability.  
	The regression and ANN rows correspond to the various statistical models including HedgeNet with three different feature sets. For these two sets of rows, the numbers are reported as relative improvements in MSHE over using the BS Delta, i.e., \eqref{eq:relative-improvement'}.
Numbers in bold represent the largest outperformance (in each column the best one is chosen along with the ones that are within 1\% of the best). 
}
	\label{tab:sp-overall}
\end{table}

\Cref{tab:sp-overall} indicates that it is easier to outperform the  BS Delta when hedging out-of-the money calls than out-of-the money puts.  However, note that the BS Delta itself reduces the MSHE more for puts than for calls when using the zero hedge as baseline. To see this, let us have a closer look at the one-day period. For calls, hedging with the BS Delta reduces the MSHE by $1-0.687/4.01\approx 83\%$, while for puts, it reduces the MSHE by  $1-0.655/4.78\approx 88\%$.  Using the Hull-White Delta reduces the MSHE for calls only by  $1-(1-0.231)\times 0.687/4.01 \approx 87\%$,
but for puts by  $1-(1-0.169)\times 0.655 / 4.78 \approx 89\%$.  Hence, the relative outperformance of the linear regressions and ANNs over the BS Delta  is higher  exactly when  the  BS Delta has a worse performance. These observations are not due to the asymmetric choice of moneyness (recall that we only consider out-of-the money options with moneyness $M = S_0/K$ between $0.8$ and $1$ for calls and between $1$ and $1.5$ for puts).  Indeed the same results as outlined in this paragraph hold true when we allow moneyness to be between $0.6$ and $1$ for calls and restrict it to be between $1$ and $1.2$ for puts. 

\begin{figure}[ht]
	\centering
\begin{subfigure}{0.48\textwidth}
\includegraphics[width=\textwidth]{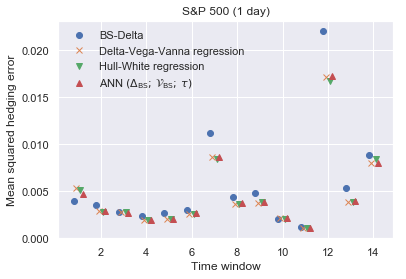}
\end{subfigure}
\quad
\begin{subfigure}{0.48\textwidth}
\includegraphics[width=\textwidth]{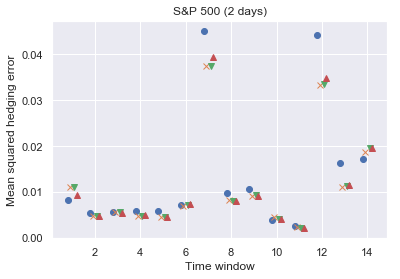}
\end{subfigure}
\caption{MSHEs of four different statistical models for the hedging ratio  across all 14 time windows in the S\&P~500 dataset, for the one-day (left) and two-day (right) hedging period. The in-sample sets for periods 7 and 12 range from 2013 to the first half of 2015 and the second half of 2015 to 2017, respectively.  The test data for the 7th time window fall exactly in the 2015-16 selloff. The test data for the 12th time window contain the first week of February 2018, where the S\&P~500 experienced a 10\% drop; see also \Cref{fig:realoptions}.}
\label{fig:sp-performance}
\end{figure}

Recall from Section~\ref{S:data} that the S\&P~500 dataset is been split in rolling windows, each time shifted by 180 days. This yields 14 out-of-sample sets. The samples in each out-of-sample set are evaluated with the model parameters estimated  on its corresponding in-sample set. \Cref{fig:sp-performance} compares the MSHEs of different statistical models by time window. Consistent with \Cref{tab:sp-overall}, the blue dots corresponding to the BS Delta are usually the largest. 
Both \Cref{tab:sp-overall} and \Cref{fig:sp-performance} show that for two-day hedging period, the MSHEs are about twice those for the one-day period.  The only exceptions are the 7th and the 13th time window, when the errors are about 4 times and 3 times larger  in the two-day period.

\begin{figure}[ht]
	\centering
	\begin{subfigure}{0.48\textwidth}
		\includegraphics[width=\textwidth]{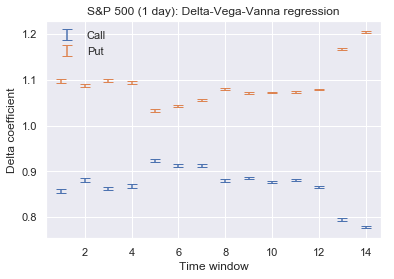}		
	\end{subfigure}
	\quad
	\begin{subfigure}{0.48\textwidth}
		\includegraphics[width=\textwidth]{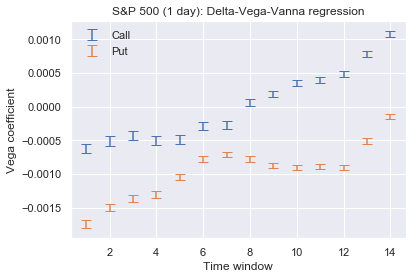}
	\end{subfigure}
	\quad
	\begin{subfigure}{0.48\textwidth}
		\includegraphics[width=\textwidth]{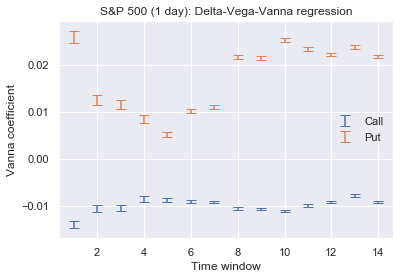}
	\end{subfigure}
	\caption{The coefficients in the Delta-Vega-Vanna regression for each of the 14 time windows in the S\&P~500 dataset. The top and bottom of each line segment are the point estimate plus/minus two standard errors. These numbers correspond to the one-day hedging period. The coefficient plots for the two-day hedging periods (not displayed here)  look very similar; in particular the Vanna coefficients for calls are again stable.  However the  Vanna  coefficients for puts and the Vega coefficients  for calls and puts are slightly more fluctuating.} 
	\label{fig:sp-dvv}
\end{figure}

\Cref{fig:sp-dvv} provides the coefficients (plus their standard errors) for the Delta-Vega-Vanna regression in the one-day period setting. The intervals are getting smaller for later time windows due to the fact that later time windows contain more samples as illustrated in Online Appendix~\ref{A:sizes}. Especially the Vanna coefficients for calls are very stable across time windows.   \Cref{fig:sp-performance} shows that both the 7th and the 12th time window, whose  out-of-sample data are the second half of 2015 and the first half of 2018, respectively, lead to an overall large MSHE. The corresponding samples are then part of the in-sample set for the following periods. And indeed, \Cref{fig:sp-dvv}  indicates a jump in some of the coefficients in the 8th and 13th time window. 
 
The Delta coefficients of calls being smaller than one implies that hedging a short position on a call, one would usually buy less of the underlying than implied by the BS Delta. On the other hand, for hedging a short position on a put, one needs to short more of the underlying. This phenomenon is consistent with  the leverage effect, discussed in Subsection~\ref{SS:Delta-vega}. Note that Vanna is positive (negative) for out-of-the money calls (puts). Hence the Vanna term in the regression further contributes to holding an even smaller number of the underlying  than  only implied by the Delta term. Since Vanna is largest in absolute value for  slightly out-of-the money options, this correction term is largest for such options. The Vega coefficients are negative for puts and most time windows also for calls, adding yet a third correction, most effective for long-dated at-the-money options.

Additional diagnostics are available in Online Appendices~\ref{A:LE} and \ref{A:addition SP500}.

We run three extra experiments to see whether the above conclusions depend on the chosen setup.
\begin{enumerate}
	\item In the first modified experiment we remove all options that have a time-to-maturity of 14 calendar days or less from both the in-sample and out-of-sample sets.
This yields an additional relative improvement of about 2\% in the one-day experiment and about 3\% in the two-day experiment for all methods presented in \Cref{tab:sp-overall}. We omit presenting the precise numbers here.
	\item In the second modified experiment we abstain from splitting the dataset in 14 time windows. Instead of 14 experiments we hence only have one, but with a much larger number of samples. We keep the ratio 4:1:1, now across the whole dataset, leading to an in-sample set of length 2850 (2280 + 570) days and a test set of length 570 days (instead of 14 test in-sample sets of length 900 (720 + 180) days and an out-of-sample set of length 180 days; see Subsection~\ref{SS:preparation}). 
	We omit the detailed results of this experiment. 
	The regression models and ANNs improve their relative performance by about 3\% to 4\% when using only one time window instead of 14 time windows.  Again the ANNs do not outperform the linear regression models. 
	\item We put the options in two roughly equally sized buckets: at-the-money/close-to-the money options and out-of-the money options. We run the linear regressions and (appropriately tuned) ANNs on both buckets separately. The bucketing tends to help the linear regressions using a single sensitivity slightly, does not change the linear regressions using several sensitivities, and leads to a worse performance of the ANNs.
\end{enumerate}
Section~\ref{S:leakage} provides a fourth experiment to check whether the cleaning process of the raw data introduced any information leakage.

\subsection{Euro Stoxx~50 tick data} \label{SS:Euro}
\Cref{tab:st-overall} shows the performance of all competing methods on the Euro Stoxx~50 dataset.  Again we conclude that the ANNs in general do not outperform the linear regressions.  
Now the Delta-Vega-Gamma-Vanna regression performs best, closely followed by the linear regressions using three sensitivities, which perform better than the Hull-White regressions.

\begin{table}[ht]
	\centering
	\begin{adjustbox}{center}
		\begin{tabular}{c c |c c c |c c c |c c c}
			& \multirow{2}{*}{} & \multicolumn{3}{c|}{1 hour} &\multicolumn{3}{c|}{1 day} & \multicolumn{3}{c}{2 days}\\
			
			& & Calls & Puts & Both & Calls & Puts & Both & Calls & Puts & Both \\
			\cmidrule{2-11}
			
			& Zero hedge & 0.431 & 1.02 & 0.756 & 4.28 & 10.2 & 7.47 & 8.20 & 24.26 & 17.4 \\
			& BS Delta & 0.109 & 0.214 & 0.167 & 1.19 & 1.99 & 1.62 & 2.97 & 4.20 & 3.67 \\
			\cmidrule{2-11}
			\multirow{3}{*}{\hspace{-1em}$\left.\begin{array}{l}
				\\
				\\
				\\
				\\
				\\
				\\
				\\
				\\
				\\
				\\
				\\
				\\
				\\
				\end{array}\text{Regressions}\right\lbrace$}
			& Delta-only & -18.9 & -11.4 & -13.6 & -21.7 & -12.2 & -15.4 & \bf-36.0 & -10.7 & -19.5 \\
			& Vega-only & \bf -25.3 & -13.8 & \bf -17.2 & \bf -23.4 & -16.0 & \bf -18.5 & -35.2 & -16.2 & -22.8 \\
			& Gamma-only & -0.62 & -1.29 & -1.10 & -15.7 & -4.64 & -8.37 & -32.7 & -4.62 & -14.4 \\
			& Vanna-only &  -16.1 & -5.18 & -8.35 & -17.1 & -12.6 & -14.1 & -26.9 & -16.9 & -20.4 \\
			& Delta-Gamma & -18.0 & -14.5 & -15.5 & -20.5 & -12.7 & -15.4 & -33.5 & -6.89 & -16.1 \\
			& Delta-Vega & -23.9 & -13.7 & -16.7 & -22.7 & -15.4 & \bf -17.9 & \bf-36.9 & -15.3 & -22.8 \\
			& Delta-Vanna & -20.8 & -11.4 & -14.1 & -19.2 & -14.8 & -16.3 & -34.9 & -17.2 & \bf -23.4 \\

			& Delta-Vega-Gamma & -21.6 & \bf -15.2 & \bf -17.0 & -20.7 & -15.4 & -17.2 & -34.4 & -13.5 & -20.8 \\
			& Delta-Vega-Vanna & -23.6 & -13.7 & -16.6 & -19.6 & -16.7 & \bf -17.7 & -35.1 & -18.5 & \bf -24.2 \\
			
			& Delta-Gamma-Vanna & -23.1 & \bf -15.5 & \bf -17.7 & -20.2 & \bf -17.9 & \bf -18.7 & -33.8 & \bf -17.7 & \bf -23.3 \\
			& Delta-Vega-Gamma-Vanna & -23.3 & \bf -15.6 & \bf -17.8 & -20.1 & \bf -18.0 & \bf -18.7 & -34.4 & \bf -18.2 & \bf -23.9 \\ 
			& Hull-White  & -20.0 & -12.5 & -14.7 & -20.7 & -14.3 & -16.4 & \bf -36.1 & -13.3 & -21.2 \\
			& Hull-White-relaxed & -20.3 & -12.6 & -14.8 & -20.6 & -14.2 & -16.4 & \bf -36.1 & -12.7 & -20.8 \\
			\cmidrule{2-11}
			\multirow{3}{*}{\hspace{1.5em}$\left.\begin{array}{l}
				\\
				\\
				\\
				\end{array}\text{ANNs}\right\lbrace$}
			& $M;\, \sigma_{\rm impl} \sqrt{\tau}$ 
			& -17.6 & \bf-15.7 & -16.3 & -8.96 & -3.3 & -5.21 & -27.4 & 11.3 & -2.12\\
			& $\Delta_{\rm BS};\, \mathcal{V}_{\rm BS};\, \tau$ & -16.1 & -6.08 & -9.01 & -19.0 & -6.83 & -10.9 & -25.6 & -3.6 & -11.2 \\
			& $\Delta_{\rm BS};\, \mathcal{V}_{\rm BS};\, 	\rm Va_{\rm BS};\, \tau$ &
			\bf-25.0 & -13.3 & -16.7 & -18.8 & -10.1 & -13.1 & -29.2 & -6.96 & -14.7
		\end{tabular}
	\end{adjustbox}
	\caption[Performance of the benchmarks and ANNs on the Euro Stoxx~50 data set, when the in-sample and out-of-sample are split into one time window.]{Performance of the benchmarks and ANNs on the Euro Stoxx~50 data set, when the in-sample and out-of-sample are split into one time window.  We refer to the caption of \Cref{tab:sp-overall} for an explanation.}
	\label{tab:st-overall}
\end{table}

 Just using the BS Delta reduces the overall MSHE by about 78\%-79\%. This percentage is very stable across the three different hedging periods and smaller than in the S\&P~500 dataset. Again, the BS Delta reduces the MSHE more for puts than for calls, and 
 the relative outperformance of the regression models is larger when the BS Delta is worse.

We list the coefficients of the Delta-Vega-Gamma-Vanna regression  (plus their standard errors)  in  \Cref{T:st coeffs}.  Again, the Delta coefficients for calls (puts) are smaller (larger) than one, consistent with  the leverage effect. Additional diagnostics are available in Online Appendices~\ref{A:LE} and \ref{A:addition Eurox50}.

	\begin{table}[ht]
		\centering
		\begin{tabular}{c | c| c | c|  c| c| c}
		 &  \multicolumn{2}{c|}{1 hour} &  \multicolumn{2}{c|}{1 day} &  \multicolumn{2}{c}{2 days} \\
		 & Calls & Puts & Calls & Puts & Calls & Puts \\
		 \cmidrule{1-7}
		 Delta & $0.944 \pm 0.002$ & $1.134 \pm 0.002$ & $0.755\pm 0.003$ & $1.056\pm 0.003$ & $0.821\pm 0.004$ & $1.021\pm 0.003$ \\
		 Vega & $-0.002\pm 0.000$ & $0.000 \pm 0.000$ & $-0.001\pm
		 0.000$ & $-0.002\pm 0.000$ & $-0.001\pm 0.000$ &$-0.002 \pm 0.000$ \\
		 Gamma & $-0.021 \pm 0.004$ & $0.213 \pm 0.003$ & $0.226\pm	 0.008$ & $0.393\pm 0.006$ & $0.109\pm 0.010$ & $0.417\pm	 0.008$\\
		 Vanna & $-0.010\pm 0.000$ & $0.014\pm 0.000$ & $0.004\pm0.000$ & $0.029\pm 0.000$ & $0.003\pm 0.000$ & $0.025\pm 0.000$\\
		\end{tabular}
	\caption{Coefficients of Delta-Vega-Gamma-Vanna regression for each sensitivity on the Euro Stoxx~50 dataset. Coefficients are presented for calls and puts separately. Each cell shows the coefficient and its standard error. }
	\label{T:st coeffs}
	\end{table}

Similarly to the S\&P~500 dataset we run two additional experiments.
\begin{enumerate}
	\item  In the first one, we only consider options with a time-to-maturity of 14 calendar days or more.
This yields an additional relative improvement of about 4\% to 8\%, in comparison with \Cref{tab:st-overall}. The improvement tends to be larger for the regressions using a smaller number of  sensitivities. In particular, the Delta-Vega-Vanna regression now seems to dominate the Delta-Vega-Gamma-Vanna regression, especially for the two-day hedging period. We again omit the precise numbers here as the overall conclusions do not change.
\item We again put the options in two roughly equally sized buckets: at-the-money/close-to-the money options and out-of-the money options. Running the statistical models on both buckets separately seems to help slightly the linear regressions with only one sensitivity but does not change or worsens the performance of the other linear regressions and ANNs.
\end{enumerate}
We also refer to Section~\ref{S:leakage} for another experiment to check how the cleaning of the data might influence the results of this subsection.

\subsection{Guidelines on statistical hedging} \label{SS:5.5}
We now develop some guidelines based on the results of the last two subsections.  

In none of the  datasets do ANNs outperform the linear regression models. We conclude that the option sensitivities suffice to capture the nonlinearities in the data that are relevant for the hedging task.  Additional drawbacks of ANNs are their computational demands and the necessary effort to tune their hyperparameters (see Online Appendix~\ref{A:reg-para}).


Next, we have a closer look at the MSHEs of the linear regression models. To this end, in the spirit of \eqref{eq:200103}, let us define the time-$t$ MSHE by 
\begin{align*}
	 \rm{MSHE}_t^\delta = \frac{1}{N_t} \sum_{j}^{N_t}   \left(100 \frac{V_{t+1, j}^\delta}{S_{t}}\right)^2,
\end{align*}
where $N_t$ denotes the number of samples at time $t$. Here, $t$ ranges over days in the test set and $\delta$ denotes one of the hedging methods. Hence $\rm{MSHE}_t^\delta$ denotes the average of a cross-section of hedging errors, namely those corresponding to the options traded at some time $t$.  Next, for each pair of hedging methods (e.g., the Delta-only and the Delta-Vega-Vanna regressions), we compute an approximate confidence interval for the difference of the MSHEs by adding and subtracting twice the standard error to the mean of the differenced time-$t$ MSHEs.  To be more specific, we denote the difference of the MSHEs between two regression models $\delta_A$ and $\delta_B$ by $\rm{MSHE}_t^{\delta_A - \delta_B} = \rm{MSHE}_t^{\delta_A} - \rm{MSHE}_t^{\delta_B}$. Then the approximate confidence interval for the two regression methods is given by
\[
\left(\frac{1}{T} \sum_{t=1}^{T}\rm{MSHE}_t^{\delta_A - \delta_B} - 2 * \text{Std}(\rm{MSHE}_t^{\delta_A - \delta_B}),\,
\frac{1}{T} \sum_{t=1}^{T}\rm{MSHE}_t^{\delta_A - \delta_B} + 2 * \text{Std}(\rm{MSHE}_t^{\delta_A - \delta_B})\right),
\]
where $T$ denotes the number of days in the test set and $\rm \mathop{std}$ denotes the (population) standard deviation.

Due to their possible statistical dependence in time, these confidence intervals need to be interpreted with caution. They allow us to make the following observations.
\begin{itemize}
\item For both hedging periods in the S\&P~500 dataset, the confidence intervals for time-$t$ MSHEs of BS Delta hedging paired with any of the statistical regressions (except for Gamma-only and Vanna-only regressions)  do not contain zero, strongly suggesting that their relative outperformance is not due to noise only. 
The same observation also holds for the one-hour and two-day hedging periods in the Euro Stoxx~50 dataset. For the one-day hedging period in the Euro Stoxx~50 dataset the statistical methods reduce the  BS Delta hedging error by up to 18.7\%, but the corresponding confidence intervals include zero. This gives an instance where the outperformance seems to be economically significant but fails to be statistically significant.
\item There is  statistical evidence for the underperformance of the Gamma-only and Vanna-only regressions. Pairing them with any of the linear regression models usually leads to confidence intervals that do not include zero. However, among any pairs of the remaining linear regression models the evidence is not clear cut. Sometimes the corresponding confidence intervals contain zero, sometimes they do not. 
\end{itemize}

We recommend to choose one of the linear regression models, for example, the Delta-Vega-Vanna or the Delta-Vega-Gamma-Vanna regressions, which perform best in the above experiments. 
Let us also note that the choice between the two probably does not matter much from an economic perspective. Indeed, let us consider the one-day hedging period in Euro Stoxx~50, where the two regressions yield a relative reduction of 17.7\% and 18.7\% (see \Cref{tab:st-overall}).  
If we now consider the Sharpe ratio of a delta-hedged option as in Subsection~\ref{SS:MSHE}, these relative reductions increase the Sharpe ratio by a factor of $1/\sqrt{0.823} \approx 1.10$ and $1/\sqrt{0.813} \approx 1.11$, respectively.  While either one leads to an economically significant increase in Sharpe ratio, their relative difference seems to be very minor.

We  conclude this section with a further observation. Motivated by the reported results we try another `fixed' hedging strategy that does not require any historical data.  All calls are hedged by $0.9*\delta_{\rm BS}$ and puts are hedged by  $1.1*\delta_{\rm BS}$.  We have not run other such `fixed' hedging strategies (hence, we have not optimised this 10\% relative correction term). 
 \Cref{tab:fixed-strategy} shows the relative performance of this `fixed'  strategy with respect to BS Delta on the S\&P~500 and Euro Stoxx~50 datasets. The out-of-sample tests are the same ones that were used for Tables~\ref{tab:sp-overall} and \ref{tab:st-overall}.
This simple strategy does very well but underperforms the linear regression models. 

\begin{table}[ht]
	\centering
	\begin{adjustbox}{center}	
		\begin{tabular}{c c |c c c |c c c |c c c}
			& \multirow{2}{*}{} & \multicolumn{3}{c|}{1 hour} &\multicolumn{3}{c|}{1 day} & \multicolumn{3}{c}{2 days}\\
			
			& & Calls & Puts & Both & Calls & Puts & Both & Calls & Puts & Both \\
			\cmidrule{2-11}
			
			& S\&P~500 & - & - & - & -18.6 & -13.1 & -14.8 & -15.0 & -11.4 & -12.6 \\
			& Euro Stoxx~50 & -15.4 & -10.3 & -11.8 & -15.4 & -12.7 & -13.6 & -23.7 & -16.6 & -19.0\\
		\end{tabular}
	\end{adjustbox}
	\caption{Performance of the `fixed' hedging strategy on the S\&P~500 and Euro Stoxx~50 datasets. In the `fixed' hedges strategy, calls (puts) are hedged by $0.9*\delta_{\rm BS}$ ($1.1*\delta_{\rm BS}$).   See the caption of \Cref{tab:sp-overall} for further explanations.}
	\label{tab:fixed-strategy}
\end{table}

\section{Potential information leakage through data cleaning} \label{S:leakage}

We next discuss information leakage issues connected to the data cleaning process.  One obvious mistake would be removing samples with \textit{wrong-way option price changes}. An example is the removal of call option samples, whenever the underlying's price increases  but the call price decreases. Although a first thought might be that this is a data issue such samples are very well possible due to changes in the bid-ask spread or due to the leverage effect; see also \cite{bakshi2000call} and \cite{perignon2006testing} for empirical  evidence.  Another important source for information leakage is introduced if the dataset is split into  in-sample and out-of-sample sets without paying respect to the time series structure. This can be mitigated by using a chronological split instead of a random split; see \cite{RW.2021.leakage}. 

The availability of end-of-period prices is a more difficult issue to be resolved. Here, in our opinion, information leakage cannot be completely avoided since it is not clear at the beginning of a period whether  prices can be observed at its end. If those prices were missing at random, it would be fine to remove those samples during backtesting. However, for financial price data, such an assumption cannot be easily justified. Indeed, missing observations tend to be caused by missing market liquidity. Market liquidity and the  implied volatility surface might  very well depend on each other. Hence, removing missing observations could potentially lead to biased parameter estimations.

To understand whether information leakage through missing price observations appears in our experiments we run robustness checks for both the S\&P~500 and the Euro Stoxx~50 datasets. 

We begin with the S\&P~500 dataset.  For these data, we have quoted prices for all options, along with trading volumes.  For the results in Subsection~\ref{SS:SP500}, we remove all samples whose trading volume at the beginning of its period are zero.  We keep those samples whose volume at the beginning is positive, but zero at the end of the period. As a robustness check we rerun the complete analysis with those samples removed whose trading volume is zero at the end of the period.  This reduces the overall dataset by about 22\% and increases the MSHE of the zero-hedge for puts (by more than 10\%). An explanation for this increase is that this modified cleaning procedure removes especially deep out-of-the-money puts, thus increasing the average squared prices changes. However, the relative performance improvement of the models with respect to the BS Delta does not change much; in particular, the conclusions of Subsection~\ref{SS:SP500} seem to be robust with respect to this cleaning procedure. 

Next, let us discuss the Euro Stoxx~50 dataset consisting of tick data. Using such tick data leads to several difficulties concerning missing price observations.  First, the underlying's prices (we use short-term futures on the Euro Stoxx~50) and option prices are not observed synchronously. This issue is relatively mild since futures are extremely liquid. For an option observation at some time $t$ we thus  use the future's price at the last transaction before $t$.  

However, a major issue in the data cleaning process is to determine the price of the option at the end of a period.  To illustrate, consider the one-hour period setup. If an option transaction in the  dataset is observed at some time $t$, then we would like to know the option price at time $t + $1 hour to backtest the hedging performance of the different methods. It is very unlikely to find a trade at exactly this time. To handle this issue we introduced a \textit{matching tolerance window} of 6 mins (see Subsection~\ref{SS:EuroStoxx}). That is, if at some time $t$ a transaction occurs then the sample's end-of-period price is the first price observation after time $t +$1 hour, and the sample is discarded if this end-of-period transaction occurs later than $t +$66 minutes. 

As discussed above, we have clearly introduced some information leakage by removing illiquid samples for which no end-of-period price is observed. Let us now do again a robustness check. To this end, we increase the matching tolerance window from 6 minutes to 30 minutes. In the one-day period situation, this increases the overall number of samples from 0.6 million to 1.4 million, a 133\% increase. This modified set contains now many more illiquid options, reflected also in a smaller MSHE of the zero-hedge.

We first summarise how the Delta-Vega-Gamma-Vanna regression performs on this modified and enlarged dataset. For the two-day hedging period, the performance improves on calls  but worsens on puts, reducing  the overall performance from about -23.9\% to -23.0\%.  For the one-day period, the longer matching tolerance window improves the Delta-Vega-Gamma-Vanna regression by $0.59\%$ with respect to BS Delta, from -18.7\% to -19.3\%, benefiting both calls and puts. For the one-hour hedging period, the overall performance worsens by 0.1\% with respect to BS Delta, from -17.8\% to -17.7\%, and the longer matching tolerance window benefits  calls and not puts.  All in all, for the regression models, the conclusions of Subsection~\ref{SS:Euro} are still valid. 
However, the longer matching tolerance window has a significantly negative effect for the ANNs. Now ANN ($\Delta_{\rm BS};\, \mathcal{V}_{\rm BS};\, 	\rm Va_{\rm BS};\, \tau$) always produce worse performance for the three hedging periods, up to even a 6\% loss in outperformance. 
Overall,  doubling the dataset by increasing the matching tolerance window does not change the regression results much, but significantly handicaps the training of the ANNs.  A further test with a matching tolerance window of 60 minutes leads to the same conclusions.

\section{Conclusion and discussion}  \label{sec:conclusion}
In this work, we consider the problem of hedging an option over one period.  We consider statistical, regression-type hedging ratios (in contrast to model-implied hedging ratios). To study whether the option sensitivities already capture the relevant nonlinearities we develop a suitable ANN architecture.
  Experiments involving both quoted prices (S\&P~500 options) and high-frequency tick data (Euro Stoxx~50 options) show that the ANNs perform roughly as well (but not better) as the sensitivity-based linear regression models. However, the ANNs are not able to find additional non-linear features. Hence option sensitivities by themselves (in particular, Delta, Vega, and Vanna) in combination with a linear regression are sufficient for a good hedging performance.

The linear regression models improve the hedging performance (in terms of MSHE) of the BS Delta by about 15-20\% in real-world datasets. An explanation is the leverage effect that allows the partial hedging of changes in the implied volatility by using the underlying.  As a rule of thumb, historical data seem to imply that calls should be hedged with about $0.9 \delta_{\rm BS}$ and puts with about $1.1 \delta_{\rm BS}$.   
With the presence of sufficient historic data we recommend to follow a hedging strategy obtained from a linear regression on the BS Delta, BS Vega, BS Vanna, and possibly the BS Gamma.

We have not performed a cross-sectional study where the hedging ratio is estimated not only from options written on the same underlying. 
It would be interesting to see whether the hedging ratios of the linear regression models can be further improved by using options written on different underlyings, e.g., the constituents of an index.

\bibliography{reference}

\begin{thebibliography}{38}
\providecommand{\natexlab}[1]{#1}
\providecommand{\url}[1]{\texttt{#1}}
\expandafter\ifx\csname urlstyle\endcsname\relax
  \providecommand{\doi}[1]{doi: #1}\else
  \providecommand{\doi}{doi: \begingroup \urlstyle{rm}\Url}\fi

\bibitem[Albrecher et~al.(2007)Albrecher, Mayer, Schoutens, and
  Tistaert]{albrecher2007little}
H.~Albrecher, P.~Mayer, W.~Schoutens, and J.~Tistaert.
\newblock The little {Heston} trap.
\newblock \emph{Wilmott}, 1:\penalty0 83--92, 2007.

\bibitem[Alexander and Nogueira(2007)]{alexander2007model}
C.~Alexander and L.~M. Nogueira.
\newblock Model-free hedge ratios and scale-invariant models.
\newblock \emph{Journal of Banking \& Finance}, 31\penalty0 (6):\penalty0
  1839--1861, 2007.

\bibitem[Alexander et~al.(2012)Alexander, Rubinov, Kalepky, and
  Leontsinis]{alexander2012regime}
C.~Alexander, A.~Rubinov, M.~Kalepky, and S.~Leontsinis.
\newblock Regime-dependent smile-adjusted delta hedging.
\newblock \emph{Journal of Futures Markets}, 32\penalty0 (3):\penalty0
  203--229, 2012.

\bibitem[Bakshi et~al.(1997)Bakshi, Cao, and Chen]{bakshi1997empirical}
G.~Bakshi, C.~Cao, and Z.~Chen.
\newblock Empirical performance of alternative option pricing models.
\newblock \emph{The Journal of Finance}, 52\penalty0 (5):\penalty0 2003--2049,
  1997.

\bibitem[Bakshi et~al.(2000)Bakshi, Cao, and Chen]{bakshi2000call}
G.~Bakshi, C.~Cao, and Z.~Chen.
\newblock Do call prices and the underlying stock always move in the same
  direction?
\newblock \emph{The Review of Financial Studies}, 13\penalty0 (3):\penalty0
  549--584, 2000.

\bibitem[Bengio(1997)]{Bengio:1997}
Y.~Bengio.
\newblock Using a financial training criterion rather than a prediction
  criterion.
\newblock \emph{International Journal of Neural Systems}, 8\penalty0
  (4):\penalty0 433--443, 1997.

\bibitem[Bergomi(2009)]{Bergomi2009}
L.~Bergomi.
\newblock Smile dynamics {IV}.
\newblock \emph{Risk}, 22:\penalty0 94--100, 2009.

\bibitem[Bertsimas et~al.(2000)Bertsimas, Kogan, and Lo]{bertsimas2000time}
D.~Bertsimas, L.~Kogan, and A.~W. Lo.
\newblock When is time continuous?
\newblock \emph{Journal of Financial Economics}, 55\penalty0 (2):\penalty0
  173--204, 2000.

\bibitem[Black(1976)]{black1976studies}
F.~Black.
\newblock Studies of stock market volatility changes.
\newblock \emph{Proceedings of the American Statistical Association Business
  and Economic Statistics Section}, 1976.

\bibitem[Boyle and Emanuel(1980)]{boyle1980discretely}
P.~P. Boyle and D.~Emanuel.
\newblock Discretely adjusted option hedges.
\newblock \emph{Journal of Financial Economics}, 8\penalty0 (3):\penalty0
  259--282, 1980.

\bibitem[Buehler et~al.(2019)Buehler, Gonon, Teichmann, and
  Wood]{buehler2019deep}
H.~Buehler, L.~Gonon, J.~Teichmann, and B.~Wood.
\newblock Deep hedging.
\newblock \emph{Quantitative Finance}, 19\penalty0 (8):\penalty0 1271--1291,
  2019.

\bibitem[Carr and Wu(2020)]{Carr.Wu.2019}
P.~Carr and L.~Wu.
\newblock Option profit and loss attribution and pricing: A new framework.
\newblock \emph{The Journal of Finance}, 75\penalty0 (4):\penalty0 2271--2316,
  2020.

\bibitem[Carverhill and Cheuk(2003)]{carverhill2003alternative}
A.~P. Carverhill and T.~H. Cheuk.
\newblock Alternative neural network approach for option pricing and hedging.
\newblock SSRN 480562, 2003.

\bibitem[Claeskens and Hjort(2003)]{Claeskens:2003}
G.~Claeskens and N.~L. Hjort.
\newblock The focused information criterion.
\newblock \emph{Journal of the American Statistical Association}, 98\penalty0
  (464):\penalty0 900--916, 2003.

\bibitem[Cont and Da~Fonseca(2002)]{cont2002dynamics}
R.~Cont and J.~Da~Fonseca.
\newblock Dynamics of implied volatility surfaces.
\newblock \emph{Quantitative Finance}, 2\penalty0 (1):\penalty0 45--60, 2002.

\bibitem[Cont and Kan(2011)]{ContKan}
R.~Cont and Y.~H. Kan.
\newblock Dynamic hedging of portfolio credit derivatives.
\newblock \emph{SIAM Journal on Financial Mathematics}, 2\penalty0
  (1):\penalty0 112--140, 2011.

\bibitem[Cr{\'e}pey(2004)]{crepey2004delta}
S.~Cr{\'e}pey.
\newblock Delta-hedging vega risk?
\newblock \emph{Quantitative Finance}, 4\penalty0 (5):\penalty0 559--579, 2004.

\bibitem[Denkl et~al.(2013)Denkl, Goy, Kallsen, Muhle-Karbe, and
  Pauwels]{denkl2013performance}
S.~Denkl, M.~Goy, J.~Kallsen, J.~Muhle-Karbe, and A.~Pauwels.
\newblock On the performance of delta hedging strategies in exponential
  {L}{\'e}vy models.
\newblock \emph{Quantitative Finance}, 13\penalty0 (8):\penalty0 1173--1184,
  2013.

\bibitem[Dugas et~al.(2009)Dugas, Bengio, B{\'e}lisle, Nadeau, and
  Garcia]{dugas2009incorporating}
C.~Dugas, Y.~Bengio, F.~B{\'e}lisle, C.~Nadeau, and R.~Garcia.
\newblock Incorporating functional knowledge in neural networks.
\newblock \emph{Journal of Machine Learning Research}, 10\penalty0
  (Jun):\penalty0 1239--1262, 2009.

\bibitem[Garcia and Gen{\c{c}}ay(2000)]{garcia2000pricing}
R.~Garcia and R.~Gen{\c{c}}ay.
\newblock Pricing and hedging derivative securities with neural networks and a
  homogeneity hint.
\newblock \emph{Journal of Econometrics}, 94\penalty0 (1-2):\penalty0 93--115,
  2000.

\bibitem[Ghysels et~al.(1998)Ghysels, Patilea, Renault, and
  Torr{\`e}s]{ghysels1997nonparametric}
E.~Ghysels, V.~Patilea, {\'E}.~Renault, and O.~Torr{\`e}s.
\newblock Nonparametric methods and option pricing.
\newblock In D.~Hand and S.~Jacka, editors, \emph{Statistics in Finance},
  chapter~13, pages 261--282. John Wiley \& Sons, 1998.

\bibitem[Glorot and Bengio(2010)]{glorot2010understanding}
X.~Glorot and Y.~Bengio.
\newblock Understanding the difficulty of training deep feedforward neural
  networks.
\newblock In \emph{Proceedings of the Thirteenth International Conference on
  Artificial Intelligence and Statistics}, pages 249--256, 2010.

\bibitem[Glorot et~al.(2011)Glorot, Bordes, and Bengio]{glorot2011deep}
X.~Glorot, A.~Bordes, and Y.~Bengio.
\newblock Deep sparse rectifier neural networks.
\newblock In \emph{Proceedings of the Fourteenth International Conference on
  Artificial Intelligence and Statistics}, pages 315--323, 2011.

\bibitem[Heston(1993)]{heston1993closed}
S.~L. Heston.
\newblock A closed-form solution for options with stochastic volatility with
  applications to bond and currency options.
\newblock \emph{The Review of Financial Studies}, 6\penalty0 (2):\penalty0
  327--343, 1993.

\bibitem[Hull and White(2017)]{hull2017optimal}
J.~Hull and A.~White.
\newblock Optimal delta hedging for options.
\newblock \emph{Journal of Banking \& Finance}, 82:\penalty0 180--190, 2017.

\bibitem[Hutchinson et~al.(1994)Hutchinson, Lo, and
  Poggio]{hutchinson1994nonparametric}
J.~M. Hutchinson, A.~W. Lo, and T.~Poggio.
\newblock A nonparametric approach to pricing and hedging derivative securities
  via learning networks.
\newblock \emph{The Journal of Finance}, 49\penalty0 (3):\penalty0 851--889,
  1994.

\bibitem[Kingma and Ba(2015)]{kingma2014adam}
D.~P. Kingma and J.~Ba.
\newblock Adam: a method for stochastic optimization.
\newblock In \emph{International Conference on Learning Representations}, 2015.

\bibitem[Krizhevsky et~al.(2012)Krizhevsky, Sutskever, and
  Hinton]{krizhevsky2012imagenet}
A.~Krizhevsky, I.~Sutskever, and G.~E. Hinton.
\newblock {ImageNet} classification with deep convolutional neural networks.
\newblock In \emph{Advances in Neural Information Processing Systems}, pages
  1097--1105, 2012.

\bibitem[Lyons(1995)]{Lyons_1995}
T.~J. Lyons.
\newblock Uncertain volatility and the risk-free synthesis of derivatives.
\newblock \emph{Applied Mathematical Finance}, 2\penalty0 (2):\penalty0
  117--133, 1995.

\bibitem[Malliaris and Salchenberger(1993)]{malliaris1993neural}
M.~Malliaris and L.~Salchenberger.
\newblock A neural network model for estimating option prices.
\newblock \emph{Journal of Applied Intelligence}, 3\penalty0 (3):\penalty0
  193--206, 1993.

\bibitem[O'Hara(1997)]{ohara1997market}
M.~O'Hara.
\newblock \emph{Market Microstructure Theory}.
\newblock Wiley, 1997.

\bibitem[P{\'e}rignon(2006)]{perignon2006testing}
C.~P{\'e}rignon.
\newblock Testing the monotonicity property of option prices.
\newblock \emph{The Journal of Derivatives}, 14\penalty0 (2):\penalty0 61--76,
  2006.

\bibitem[Ruf and Wang(2020)]{RW.2019.LR}
J.~Ruf and W.~Wang.
\newblock Neural networks for option pricing and hedging: {a} literature
  review.
\newblock \emph{Journal of Computational Finance}, 24\penalty0 (1):\penalty0
  1--46, 2020.

\bibitem[Ruf and Wang(2021)]{RW.2021.leakage}
J.~Ruf and W.~Wang.
\newblock Information leakage in backtesting.
\newblock SSRN 3836631, 2021.

\bibitem[Stoll(1978)]{stoll1978supply}
H.~R. Stoll.
\newblock The supply of dealer services in securities markets.
\newblock \emph{The Journal of Finance}, 33\penalty0 (4):\penalty0 1133--1151,
  1978.

\bibitem[Tankov and Voltchkova(2009)]{tankov2009asymptotic}
P.~Tankov and E.~Voltchkova.
\newblock Asymptotic analysis of hedging errors in models with jumps.
\newblock \emph{Stochastic Processes and Their Applications}, 119\penalty0
  (6):\penalty0 2004--2027, 2009.

\bibitem[V{\"a}h{\"a}maa(2004)]{vahamaa2004delta}
S.~V{\"a}h{\"a}maa.
\newblock Delta hedging with the smile.
\newblock \emph{Financial Markets and Portfolio Management}, 18\penalty0
  (3):\penalty0 241--255, 2004.

\bibitem[Yarotsky(2017)]{yarotsky2017error}
D.~Yarotsky.
\newblock Error bounds for approximations with deep {ReLU} networks.
\newblock \emph{Neural Networks}, 94:\penalty0 103--114, 2017.

\end{thebibliography}

\begin{appendices}

In this online appendix we provide additional material. Appendix~\ref{A:simulation} provides a simulation study, where the data-generating mechanism is known.
Appendix~\ref{A:reg-para} provides additional details on the training of the ANNs. 
 Appendix~\ref{A:sizes} gives details on the cleaning of the S\&P~500 and Euro Stoxx~50 datasets and presents some descriptive statistics. Appendix~\ref{A:LE} looks closer at the leverage effect in the  S\&P~500 and Euro Stoxx~50 datasets. Appendices~\ref{A:addition SP500} and \ref{A:addition Eurox50} provide more diagnostics on the empirical results presented in Section~\ref{S:Results}.

\section{Simulation study} \label{A:simulation}
This online appendix presents an extensive simulation study to compare the performance of the ANNs and the linear regression models when the data are generated from a known model; here,  the Black-Scholes and the Heston models.
Subsection~\ref{SS:BS&H} explains the data-generating mechanism for the simulated datasets.
Subsections~\ref{SS:BS} and \ref{SS:Heston} provide and discuss the results for this data.

\subsection{Simulation of the Black-Scholes and Heston datasets} \label{SS:BS&H}
For the simulation study  two data-generating mechanisms are considered. In the first one, the underlying's price process is simulated from the Black-Scholes
 stochastic integral equation
\begin{equation*}
\label{eq:GBM}
S_t = 2000 + \mu \int_{0}^{t} S_u \mathrm{d}u + \sigma \int_{0}^{t} S_u \mathrm{d}W_u,
\end{equation*}
with  annualised rate of return $\mu = 0.1$, and annualised volatility $\sigma = 0.2$. 
In the second example the underlying's price process is simulated from the \cite{heston1993closed} model given by 
\begin{align*}
S_t &= 2000 +  \int_{0}^{t} \sqrt{Y_u} S_u \mathrm{d}W_u;\\
Y_t &= Y_0 + \kappa \int_{0}^{t}  \left(\theta-Y_u\right) \mathrm{d} u + \sigma_Y \int_0^t  \sqrt{Y_{u}} \mathrm{d} \widetilde{W}_{u};\\
\rm{Cov}(W_t, \widetilde W_t) &= \rho t,
\end{align*} 
with initial and long-term variance  $Y_0 = \theta = 0.04$, rate of mean reversion $\kappa = 5$, volatility of variance $\sigma_Y = 0.3$, and correlation $\rho = -0.6$.
Here the volatility $\sqrt{Y_t}$ of the underlying is stochastic and modelled as the square root of a process mean-reverting to $0.04$. Thanks to Feller's test of explosions, the volatility is always strictly positive.  We intentionally omit the drift to focus on the role that stochastic volatility plays.

We first simulate $1.25$ years of the underlying's price from the Black-Scholes and Heston model, respectively. For the Black-Scholes dataset, we use exact simulation. For the Heston dataset  we use a standard Euler and  Milstein scheme. The initial value of 2000 is relevant to get a realistic number of options as their generation depends on the underlying's absolute value, as we explain next.

Along the simulated spot path, options are created following the Chicago Board Options Exchange (CBOE) rules. Details of these rules are provided on \url{http://www.cboe.com/products/stock-index-options-spx-rut-msci-ftse/s-p-500-index-options/s-p-500-options-with-a-m-settlement-spx/spx-options-specs}. The idea is the following. The option expiration date is always the fourth Friday of its expiration month. The expiration months are the 12 immediate calendar months, plus some additional long-term months (we do not generate options for those long-term months). At each expiration date new options are created, so that the market still trades options with 12 expiration months. In general, the strike price step is set to $5$ Dollars. The two strike prices closest to the current underlying's price are initially listed. If the underlying's price is  close to any one of the two strikes, a third strike will be included to cover the larger range. New series are generally added when the underlying's price trades through the highest or lowest available strike price for each expiration.

Next,  we price the options on each trading day using the Black-Scholes formula and the standard pricing formulas available for Heston, respectively;  see, for example, \cite{albrecher2007little}. Here, we set the dividend and interest rate to zero.  Moreover, in the Heston case, we fix this pricing  measure, under which $\widetilde W$ is also Brownian motion.

These 1.25 years of simulated data correspond to the in-sample data (training and validation), on which the benchmarks and ANNs are trained. 
To estimate the MSHE,  more data are simulated; however those data are only used to estimate the out-of-sample performance of the different statistical models. Choosing a time length of $1.25$ years is done for the following reason.  As explained  in Subsection~\ref{SS:preparation}, when training the ANNs for the real-world datasets, we split the data up in training, validation, and out-of-sample (test) data using the ratio  4:1:1. For the simulated datasets we keep this ratio and choose the training set to be one year long. This then yields $1.25$ years of training and validation data.  Simulating options according to the CBOE rules yields roughly the same magnitude of training data as available in each time window of the real-world datasets.

After computing the option prices and the sensitivities necessary for the statistical models,  the data are again arranged so that each row corresponds to exactly one option at one day. 
Finally, samples with option price less than $0.01$ (the tick size) or moneyness $M$ outside of the interval $[0.8, 1.5]$ are removed.  This means that if an option  has a time-to-maturity of  90 trading days, it might appear, for example, 85 times in the dataset. The option might have a moneyness outside of the interval or a too small price  for the other four trading days.

\begin{figure}[ht]
	\centering
	\includegraphics[width=0.6\textwidth]{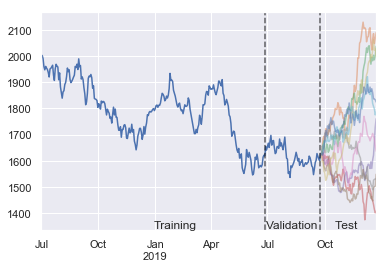}
	\caption{The single simulated price path on which options are created for the in-sample set, and  the multiple  paths on which options  are created for the out-of-sample sets. To improe the estimate of the out-of-sample MSHE, we compare the different methods below for each out-of-sample set and average them.}
	\label{fig:simulation_paths}
\end{figure}

The Black-Scholes and Heston datasets consist both of a single time window of 1.5 years. The first 450 days form the in-sample set. For the ANN, the 450 days are furthermore split into 360 (training) and 90 (validation) days. To get a more precise estimate of the MSHE, twenty out-of-sample sets of 90 days are simulated, as illustrated in \Cref{fig:simulation_paths}. 

We conclude by summarising that the in-sample dataset in the Black-Scholes dataset is 0.36 million and in the Heston dataset  0.26 million.   
 As explained above,  we created options according to the CBOE rules and then removed all  in-the-money samples. Since the underlying tends to move upwards in the Black-Scholes dataset (the drift rate was set to 10\%) we expect to have more out-of-the money put samples than call samples.  Indeed, an investigation of the Black-Scholes  dataset yields that we have roughly 91k  call samples and 277k  put samples in the in-sample set. It turns out that the Heston in-sample dataset, just by chance (the simulated underlying's path process moves from 2000 to about 2600) also has more put samples (192k) than call samples (69k).

\subsection{Results: Black-Scholes model} \label{SS:BS}
As reported in \Cref{tab:bs-overall}, in the one-day hedging period, the BS Delta performs best (with the exception of the Vanna-only and Vega-only regressions).  For the two-day hedging period, all regressions outperform the BS Delta. Relative to the BS Delta the regressions are about 2\% to 3\% better. 
 At first glance, this seems surprising since the BS Delta should be  close to optimal for data generated from the Black-Scholes model. Indeed, in both hedging periods, using  the BS Delta instead of not hedging at all reduces the MSHE by about 99\%. 
 
 What is happening?  Recall that we do not hedge continuously but only once in each hedging period. 
During the hedging period, the underlying's price changes, and thus, the  BS Delta chosen at the beginning of the hedging period is not optimal at other times during the hedging period.  
Since the underlying's price path has been simulated with an annualised drift rate of 10\% (see Subsection~\ref{SS:BS&H}), in average the option's Delta tends to increase over the hedging period. 
The linear regressions are able to capture this effect. For example, in the Delta-only regression, the Delta coefficient is slightly larger than one for out-of-the money calls and slightly smaller than one for out-of-the money puts (in which case the BS Delta is negative). This is in line with the observation that  the option's Delta increases over the hedging period in average.

For the one-day hedging period this drift effect is not strong enough for the linear regression models to outperform; they tend to slightly overfit to the in-sample data. For the two-day hedging period, however, this drift effect is captured by the linear regressions, as can be seen in \Cref{tab:bs-overall}. The ANNs are not able to capture this effect, due to overfitting. 

We have run another experiment, where we set the drift rate of the underlying's price path to zero and leave all others parameters the same.  In this case, the linear regressions underperform (overperform)  relative to the BS Delta by about 0.5\% for the one-day (two-day) hedging period.  Again, ANNs have the lowest performance among all considered models.

\begin{table}[ht]
	
	\begin{tabular}{c c |c c c |c c c}
		& \multirow{2}{*}{} & \multicolumn{3}{c|}{1 day} & \multicolumn{3}{c}{2 days}\\
		
		& & Calls & Puts & Both & Calls & Puts & Both \\
		\cmidrule{2-8}
		
		& Zero hedge & 27.0 & 12.3 & 16.0 & 54.9 & 23.4 & 31.2\\
		& BS Delta  & 0.164 & \bf 0.094 & 0.111 & 0.719 & 0.341 & 0.437 \\
		\cmidrule{2-8}
		\multirow{3}{*}{\hspace{-1em}$\left.\begin{array}{l}
			\\
			\\
			\\
			\\
			\\
			\\
			\\
			\\
			\\
			\\
			\\
			\\
			\\
			\end{array}\text{Regressions}\right\lbrace$}
		& Delta-only &  \bf -1.38 & 1.05 & 0.11 & -4.74 & -0.82 & -2.12\\
		& Gamma-only & \bf -1.25 &  0.97 & 0.12 & \bf -6.27 & \bf -1.39 & \bf -2.76 \\
		& Vega-only & \bf -1.22 &  0.81 & \bf -0.02 & -3.85 & -0.56 & -1.68 \\
		& Vanna-only & \bf -1.64 &  0.32 & \bf -0.46 & -5.61 & -0.60 & -1.99\\
		& Delta-Gamma & \bf -1.38 &  0.96 & 0.07 & \bf -6.26 & \bf -1.42 & \bf -2.79 \\
		& Delta-Vega &  -1.09 &  1.1 & 0.35 & -4.97 & -0.89 & \bf -2.27 \\
		& Delta-Vanna & \bf -1.30 &  1.01 & 0.11 & \bf -6.03 & -0.83 & \bf -2.47 \\
		
		& Delta-Vega-Gamma & \bf -1.16 &  0.99 & 0.21 & \bf -6.37 & \bf -1.28 & \bf -2.78 \\
		& Delta-Vega-Vanna & \bf -1.30 &  1.08 & 0.24 & \bf -6.49 & \bf -1.06 & \bf -2.68 \\
		& Delta-Gamma-Vanna & -0.85 &  0.99 & 0.31 & \bf -6.6 & \bf -1.2 & \bf -2.85 \\
		& Delta-Vega-Gamma-Vanna & -1.03 &  0.98 & 0.26 & \bf -6.62 & \bf -1.2 & \bf -2.86 \\ 
		& Hull-White  & \bf -1.44 &  1.02 & 0.07 & \bf -6.26 & -0.77 & \bf -2.46 \\
		& Relaxed Hull-White & \bf-1.43 & 1.02 & 0.07 & \bf-6.24 & -0.77 & \bf-2.45\\
		\cmidrule{2-8}
		\multirow{2}{*}{\hspace{1.5em}$\left.\begin{array}{l}
			\\
			\\
			\\
			\end{array}\text{ANNs}\right\lbrace$}
		& $M;\, \sigma_{\rm impl} \sqrt{\tau}$ & 8.9 & 2.55 & 5.65 & -3.21 & 0.55 & 0.08 \\
		& $\Delta_{\rm BS};\, \mathcal{V}_{\rm BS};\, \tau$ & 2.11 & 2.81 & 2.16 & -5.37 & 5.45 & 2.63\\
		& $\Delta_{\rm BS};\, \mathcal{V}_{\rm BS};\, {\rm Va}_{\rm BS};\, \tau$
		& -0.16 & 1.07 & 1.37 & -5.83 & 2.44& -0.21 \\
		
	\end{tabular}
	\caption{Performance of the benchmarks and ANNs on the Black-Scholes simulated dataset.  See the caption of \Cref{tab:sp-overall} for further explanations.}
	\label{tab:bs-overall}
\end{table}

We conclude this subsection with a remark. The experiments above are done with a realistic amount of samples in the in-sample set, namely obtained by following the CBOE rules on generating options as outlined in the previous subsection.  If the in-sample set was to be augmented by additional data then eventually the overfitting of the statistical models in the one-day hedging period would disappear. Moreover,  the more complex models will then outperform the simpler ones in the horse race of \Cref{tab:bs-overall}.

\subsection{Results:  Heston model} \label{SS:Heston}
For the Heston dataset, we report the numbers in \Cref{T:Heston}. Again the ANNs do not lead to a better performance than the regression models. Using the BS Delta reduces the variance by more than 97\% (96\%) for both calls and puts,  for the one-day  (two-day) hedging period. This is a larger improvement than for the real-world datasets. Note that we have roughly 3 times more put samples than call samples in the in-sample test as Subsection~\ref{SS:BS&H} explains. The coefficients for the Delta-only regression for the one-day (two-day) hedging period are for calls 0.97 (0.99) and for   puts 1.03 (1.03), all with standard deviation $\pm$ 0.001 or less. 

Note the consistently worse relative performance for hedging calls than for hedging puts in the two-day hedging period in \Cref{T:Heston}.   The BS Delta itself already performs better for calls than for puts; hence it is more difficult to improve on it  in the case of calls than for puts. 
Indeed, there are two effects in play. They cancel each other for calls but reinforce themselves for puts. (a) For out-of-the money puts and calls convexity together with time-discrete hedging suggests a  larger hedging ratio (in absolute terms).  (b) The leverage effect suggests a lower hedging ratio for calls but a larger  hedging ratio (in absolute terms) for puts. Since these two effects for calls go in opposite directions, but not for puts, the BS Delta performs better for calls than for puts.

\begin{table}[ht]
	
	\begin{tabular}{c c |c c c |c c c}
		& \multirow{2}{*}{} & \multicolumn{3}{c|}{1 day} & \multicolumn{3}{c}{2 days}\\
		
		& & Calls & Puts & Both & Calls & Puts & Both \\
		\cmidrule{2-8}
		
		& Zero hedge & 21.7 & 14.7 & 16.5 & 45.6 & 32.0 & 34.3\\
		& BS Delta  & 0.637 & 0.505 & 0.526 & 1.61 & 1.36 & 1.35\\
		\cmidrule{2-8}
		\multirow{3}{*}{\hspace{-1.25em}$\left.\begin{array}{l}
			\\
			\\
			\\
			\\
			\\
			\\
			\\
			\\
			\\
			\\
			\\
			\\
			\\
			\end{array}\text{Regressions}\right\lbrace$}
		& Delta-only & \bf-3.73 & -4.80 & -4.50 & \bf-1.09 & -4.86 & \bf-2.59\\
		& Gamma-only & -3.44 & -4.55 & -4.39 & -0.74 & -4.97 & -2.33\\
		& Vega-only & -3.20 & -3.77 & -3.77 & \bf-1.21 & -3.97 & -2.34\\
		& Vanna-only & -3.38 & -2.97 & -3.62 & \bf-1.46 & -3.54 & -2.30\\
		& Delta-Gamma & \bf-3.98 & -5.02 & \bf-4.82 & -0.92 & \bf-5.04 & -2.47\\
		& Delta-Vega & -3.51 & -4.84 & -4.39 & -0.97 & -3.89 & -2.03\\
		& Delta-Vanna & \bf-4.04 & -5.14 & \bf-4.92 & \bf-1.53 & \bf-5.42 & \bf-3.03\\

		& Delta-Vega-Gamma & \bf-3.64 & -4.97 & -4.67 & \bf-1.06 & -4.37 & -2.25\\
		& Delta-Vega-Vanna & \bf-4.07 & \bf-5.36 & \bf-5.03 & \bf-1.23 & -4.74 & -2.46\\
		& Delta-Gamma-Vanna & \bf-3.97 & -4.92 & \bf-4.77 & \bf-1.43 & -4.62 & -2.56\\
		& Delta-Vega-Gamma-Vanna & \bf-4.13 & -5.22 & \bf-4.96 & \bf-1.26 & -4.62 & -2.43\\ 
		& Hull-White  & \bf-4.12 & -5.02 & \bf-4.92 & \bf-1.23 & \bf-5.15 & \bf-2.75\\
		
		& Relaxed Hull-White & \bf-4.11 & \bf-5.02 & \bf-4.92 & \bf-1.21 & -5.16 & \bf-2.74\\
		
		\cmidrule{2-8}
		\multirow{2}{*}{\hspace{1em}$\left.\begin{array}{l}
			\\
			\\
			\\
			\end{array}\text{ANNs}\right\lbrace$}
		& $M;\, \sigma_{\rm impl} \sqrt{\tau}$ 
		& 4.49 & \bf-5.49 & 1.36 & 6.04 & -5.01 & 2.96\\
		& $\Delta_{\rm BS};\, \mathcal{V}_{\rm BS};\, \tau$ 
		& -3.01 & -5.08 & \bf-4.13 & 0.74 & -3.46 & 0.19\\
		& $\Delta_{\rm BS};\, \mathcal{V}_{\rm BS};\, {\rm Va}_{\rm BS};\, \tau$
		& -2.46 & \bf-5.68 & -3.77 & -0.27 & -2.05 & 0.01 \\
	\end{tabular}
	\caption{Performance of the benchmarks and ANNs on the Heston dataset.  See the caption of \Cref{tab:sp-overall} for further explanations.} \label{T:Heston}
\end{table}

The same remark as at the end of the previous subsection also applies here. In additional experiments, we have augmented the data with additionally simulated samples.  Eventually, the more complex models always outperform the simpler ones.  The results as displayed  in \Cref{T:Heston} show that with a limited amount of data sometimes simpler models outperform more complex ones.

As a sanity check, we consider two model-implied hedging strategies on the one-day period. The first one relies on $\delta_{\rm HS}$, the sensitivity of the option price with respect to the underlying price, computed under the Heston model with the correct parameters. This sensitivity is then adjusted by a multiple of $\nu_{\rm HS}$,  
the sensitivity of the option price with respect to the underlying variance $Y_0$. More precisely, the hedging strategy is given by
\begin{equation*}
\delta_{\rm HS} + \nu_{\rm HS} \frac{\rho \sigma_Y}{S_0};
\end{equation*}
see, for example, \cite{alexander2007model} for a derivation via a Taylor series expansion. Using this strategy leads to a reduction of $6.18\%$  (calls only: $4.99\%$; puts only: $6.73\%$) of the MSHE relative to using the BS Delta.  We note that the Delta-Vega-Vanna regression, which does not use any model information and leads to a reduction of $5.03\%$, performs almost as well as this model-specific hedging strategy.

The second model-implied hedging strategy,  suggested  by \cite{bakshi1997empirical}, is often called `Delta-Vega-neutral strategy.' It differs from all other hedging strategies used in this paper, in so far that it 
uses a second hedging instrument; here an at-the-money (ATM) call with maturity equal to one month. 
The number $\eta$ of at-the-money options held is chosen to satisfy 
\begin{align} \label{eq:210503}
	\eta  \nu^{\rm ATM}_{\rm HS} - \nu_{\rm HS} = 0, 
\end{align}
i.e., to cancel out the `Vega' risk in the hedged portfolio.  The number of stocks held is then set equal to 
$\delta_{\rm HS} - \eta  \delta^{\rm ATM}_{\rm HS} $. Relative to using the BS Delta, this hedging strategy leads to a reduction of the MSHE by  $63.8\%$ (calls only: $62.6\%$; puts only: $69.2\%$). None of the hedging strategies discussed above gets close to this one.  

We conclude this section with a brief discussion of this article's statistical hedging methods in the presence of two hedging instruments. As in the previous paragraph the hedging ratio to be determined is described as a pair $(\delta, \eta)$.  HedgeNet, i.e., the ANN architecture in Subsection~\ref{SS:architecture}, can be easily adapted to this situation: the trainable part now returns the pair $(\delta, \eta)$, and the replication value equals
\[
\widehat C_1= \delta S_1 + \eta C_1^{\rm ATM} + (1+r_{\rm onr} \Delta t) (C_0 - \delta S_0 - \eta C_0^{\rm ATM}).
\]
The formulation of a linear regression model as a benchmark seems to require more effort. A first ansatz could be to specify $\eta$ as in \eqref{eq:210503}, but with the sensitivities computed in the Black-Scholes model. 
Then $\delta$ can be written again as a linear function, but whose input now are the sensitivities of a portfolio that consists of the option to be hedged and $\eta$ ATM options. We leave the question of how well such an ANN and a linear regression work, and how to specify more flexible regression models, to future research.

\section{Additional hyperparameters of HedgeNet} \label{A:reg-para}
In this section we add details on the implementation and training of HedgeNet (see Subsection~\ref{SS:architecture}).

 Based on preliminary experiments on simulated data we set the learning rate to $10^{-4}$ and the batch size to 64. Usually we train each ANN for $300$ epochs. (We also apply visual inspections of the training / validation loss to confirm that the ANN is indeed trained.) Using a validation set, we  apply \textit{early stopping} by choosing the ANN with the smallest validation error. 

The optimisation criterion is a Tikhonov regularised version of squared loss. We use an $L^2$ penalty term for the ANN weights. We also have experimented with other regularisations, such as an $L^1$ penalty, a combined $L^1$-$L^2$ penalty, and dropout. They all lead to similar results and the same conclusions. 
The regularisation strength $\alpha$ is tuned for each dataset and hedging period. The larger $\alpha$ is the more the weights are pushed to zero. In case of the simulated data, $\alpha$ is tuned by using an independent dataset that is simulated from the same model but with a different random seed. Hence, the actual training and test datasets are different from the ones used for tuning.
For the S\&P~500 dataset, we tune only using the first four  time windows.   For the Euro Stoxx~50 dataset,  due to its single window experimental setup, we choose the tuning parameter $\alpha$ on the validation set. Such in-sample tuning favours the performance of the ANNs.  With a proper out-of-sample tuning, the ANNs would perform worse.

For each dataset and each value $\alpha$ on a logarithmic grid, we run five iterations of the ANN training, each with a different (random) weight initialisation.  For each dataset we then pick an $\alpha$ after inspecting the average and standard deviations of the test errors (on the independent dataset when using simulated data,  on the first four time windows when using the S\&P~500 dataset, and on the whole validation set when using the Euro Stoxx~50 dataset).
\Cref{tab:reg-alpha} summarises the chosen $L^2$ regularisation parameters.

	\begin{table}[ht]
	\centering
	\begin{tabular}{c  c | c c c c}
		& & S\&P~500 & Euro Stoxx~50 & Black-Scholes & Heston \\
		\cmidrule{1-6}
		\multirow{3}{*}{$M;\, \sigma_{\rm impl} \sqrt{\tau}$} & 1H & - & $10^{-5}$ & - & - \\
		& 1D & $10^{-7}$ & $10^{-2}$ & $10^{-4}$ & $10^{-4}$\\
		& 2D & $10^{-3}$ & $10^{-2}$ & $10^{-4}$ & $10^{-4}$ \\
		\cmidrule{1-6}
		\multirow{3}{*}{$\Delta_{\rm BS};\, \mathcal{V}_{\rm BS};\, \tau$} &  
		1H & -& $10^{-3}$ & - & - \\
		& 1D & $10^{-4}$ & $10^{-2}$ & $10^{-4}$ & $10^{-3}$\\
		& 2D & $10^{-3}$ & $10^{-1}$ & $10^{-3}$ & $10^{-3}$\\
		\cmidrule{1-6} 
		\multirow{3}{*}{$\Delta_{\rm BS};\, \mathcal{V}_{\rm BS};\, {\rm Va}_{\rm BS};\, \tau$}
		& 1H & - & $10^{-3}$ & - & -\\
		& 1D & $10^{-4}$ & $10^{-3}$ & $10^{-4}$ & $10^{-3}$\\
		& 2D & $10^{-2}$ & $10^{-3}$ & $10^{-3}$ & $10^{-3}$\\
	\end{tabular}
	\caption{Regularisation parameters used for the training of HedgeNet in the different experiments.}
	\label{tab:reg-alpha}
\end{table}

\section{Further details on the S\&P~500 and Euro Stoxx~50 datasets} \label{A:sizes}
We first review how we cleaned the two datasets. Then, in Subsection~\ref{A:sizes_subsection}, we provide some summary statistics concerning the corresponding in-sample and out-of-sample sets. 

\subsection{Cleaning of the datasets} \label{A:cleaning}
In the cleaning process of the S\&P~500 dataset, we remove the following samples:
\begin{itemize}
	\itemsep0em
	\item Samples with negative time value. 
	\item Samples with time-to-maturity less than 1 day.
	\item Samples where the moneyness is outside the interval $[0.8, 1.5]$. 
	\item Samples with an implied volatility higher than 100\% or smaller than 1\%.
	\item Samples with zero trading volume.
	\item Samples where the ask  is at least twice the bid.
	\item Samples with bid less than $0.05$. 
	\item Samples that do not have available next trade prices.
\end{itemize}

In the cleaning process of the Euro Stoxx~50 dataset, we remove the following samples:
\begin{itemize}
	\itemsep0em
	\item Samples with negative time value.
	\item Samples with time-to-maturity less than 1 day.
	\item Samples where the moneyness is outside the interval $[0.8, 1.5]$. 
	\item Samples with an implied volatility higher than 100\% or smaller than 1\%.
	\item Samples on expiry dates of a future. 
	\item Samples that cannot be matched to a next trade (within the matching tolerance window of 6 minutes).
	\item Samples that are traded in the first or last half an hour of each trading day.
\end{itemize}

We have run several  checks with different cleaning procedures (e.g., not removing samples with very large implied volatility, of which there are only a few in both datasets). The results of the paper are robust with respect to these modifications.

\subsection{Sizes of in-sample and out-of-sample sets} \label{A:sizes_subsection}
\begin{figure}[ht]
	\centering
	\begin{subfigure}{0.48\textwidth}
		\includegraphics[width=\textwidth]{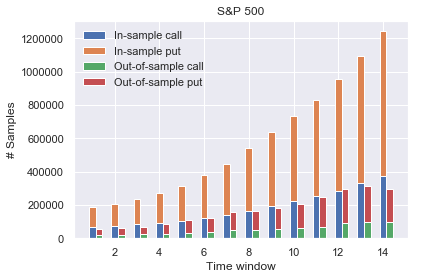}	
	\end{subfigure}
	\quad
	\begin{subfigure}{0.48\textwidth}
		\includegraphics[width=\textwidth]{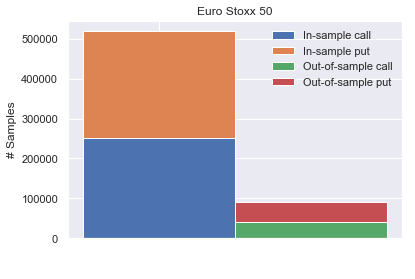}
	\end{subfigure}
	\caption{Sample size of out-of-the-money and at-the-money calls and puts in in-sample and out-of-sample sets. The left panel corresponds to the S\&P~500 dataset,  the right panel to the Euro Stoxx~50 dataset.} 
	\label{fig:num-samples}
\end{figure}

Recall that we only consider out-of-the-money  and at-the-money options.
\Cref{fig:num-samples} shows the number of samples  in each time window for the S\&P~500 and Euro Stoxx~50 datasets. 
For the S\&P~500 data (ranging from 2010 to 2019), the overall number of samples  is  2.6 million. On average, there are  1144 samples per trading day. In each time window,  the number of total samples grows continually. More puts than calls are traded, and the number of puts traded grows faster than that of calls traded. For the Euro Stoxx~50 data (ranging from 2016 to 2018),  the number of samples overall is 0.62 million.  On average, there are 988 samples per trading day. Roughly the same number of puts and calls are traded.

\begin{figure}[ht]
	\begin{subfigure}{0.48\textwidth}
		\includegraphics[width=\textwidth]{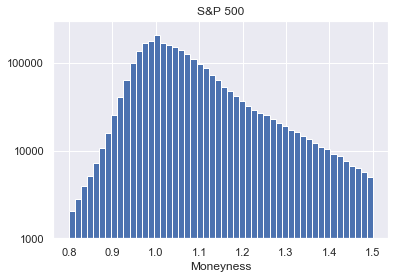}
	\end{subfigure}
	\quad	
	\begin{subfigure}{0.48\textwidth}
		\includegraphics[width=\textwidth]{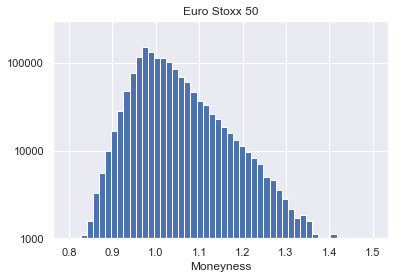}
	\end{subfigure}	
	\caption{Histogram of moneyness in the S\&P~500 (left panel) and the Euro Stoxx~50 (right panel) datasets. Samples with moneyness less than 1 correspond to calls, and samples with moneyness greater than 1 to puts.}
	\label{fig:sp-st-m-distribution}
\end{figure}

 \Cref{fig:sp-st-m-distribution} shows the distribution of moneyness in the S\&P~500 and Euro Stoxx~50 datasets. As we only consider  out-of-the-money and at-the-money options each sample with moneyness less than 1 corresponds to a call, and similarly,  each sample with moneyness greater  than 1 corresponds to a put. The distribution of moneyness for Euro Stoxx~50 data is more concentrated around a moneyness of 1. This difference is explained by the fact that the  S\&P~500 dataset consists of end-of-day quotations of all listed options, while the Euro Stoxx~50 dataset consists of tick prices of all traded options. Since close-to-the money options are more frequently traded, the  Euro Stoxx~50 dataset hence has relatively more such samples.
 
\begin{figure}[ht]
	\begin{subfigure}{0.48\textwidth}
		\includegraphics[width=\textwidth]{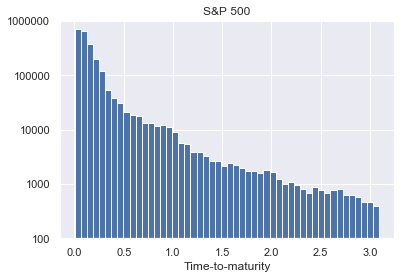}
	\end{subfigure}
	\quad	
	\begin{subfigure}{0.48\textwidth}
		\includegraphics[width=\textwidth]{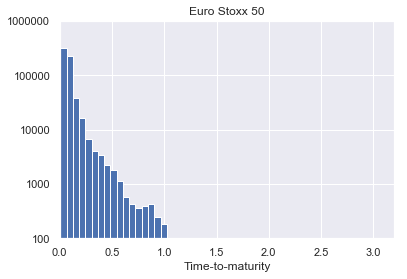}
	\end{subfigure}	
	\caption{Histogram of time-to-maturity in the S\&P~500 (left panel) and Euro Stoxx~50 (right panel) datasets.}
	\label{fig:sp-st-tau-distribution}
\end{figure}

 \Cref{fig:sp-st-tau-distribution} shows the distribution of time-to-maturity for both datasets. The S\&P~500 dataset has many more long-dated options than the Euro Stoxx~50 dataset.

\section{Some heuristics on the leverage effect} \label{A:LE}
To understand the leverage effect and its interaction with the coefficients of the linear regressions a bit better we perform the following empirical study.  For each option type (put or call) and for different time-to-maturities (namely $\tau$ smaller than 1 month, $\tau$ between 1 and 6 months, and $\tau$ greater than 6 months) we regress $\Delta \sigma_{\rm impl}$ on $\Delta S$, without intercept.  This yields a slope $b$. We then compute 
\begin{align} \label{eq:200321}
 	\rm{LC} = b\, \frac{1}{\rm N_{train}} \sum_{\rm t, j}^{\rm N_{train}} \frac{\mathcal{V}_{\rm BS,  t, j}}{\delta_{\rm BS, t, j}},
\end{align}
which we call {\em leverage coefficient}.  These heuristics are motivated by how much we should adjust a hedge due to the leverage effect. Indeed, a change of $\Delta \sigma_{\rm impl}$ leads roughly to a change of $\mathcal{V}_{\rm BS} \Delta \sigma_{\rm impl}$ in the option price. A part  $\mathcal{V}_{\rm BS} b  \Delta S$ of this change can be explained by the change in the underlying's price due to the correlation of implied volatilities and returns. Considering a multiplicative effect on the BS Delta, we need to divide this number by $\delta_{\rm BS}$.

\begin{figure}[ht]
	\centering
	\begin{subfigure}{0.48\textwidth}
		\includegraphics[width=\textwidth]{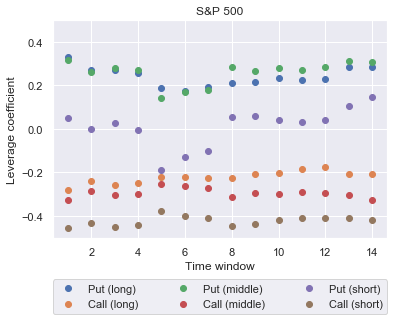}
	\end{subfigure}
	\quad
	\begin{subfigure}{0.48\textwidth}
		\includegraphics[width=\textwidth]{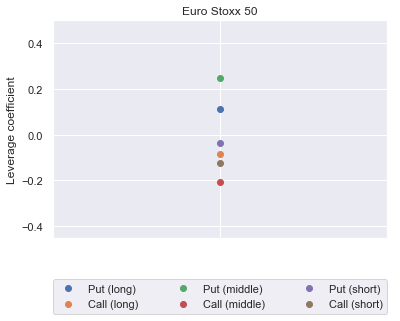}
	\end{subfigure}
	
	\caption{Leverage coefficients as given in \eqref{eq:200321} on the three categories of time-to-maturity in the S\&P~500 (left) and Euro Stoxx~50 (right) dataset for the one-day hedging period. `Short' means a time-to-maturity of less than 1 month, `middle'  means between 1 month and 6 months, and `long' means more than 6 months.}
	\label{fig:leverage-by-maturity}
\end{figure}

\Cref{fig:leverage-by-maturity} shows the leverage coefficients for the different  option categories for the one-day hedging period. The plots for the other hedging periods (for which $\Delta \sigma_{\rm impl}$ and $\Delta S$ are different, yielding slightly different estimates for $b$ in \eqref{eq:200321}) look  similar.  The fact that the leverage coefficient tends to be negative for calls (positive for puts) reflects how the regression models replace the BS Delta by a number smaller (larger) than one.
Note the jumps of the leverage coefficient in the S\&P~500 plot from period 4 to 5, 7 to 8, and 12 to 13. This is consistent with the change of the Delta coefficient in Delta-Vega-Vanna regression of \Cref{fig:sp-dvv}.

\section{Additional diagnostics for the S\&P~500 dataset} \label{A:addition SP500}
We use this appendix to provide some additional figures concerning the performance of the various statistical models on the S\&P~500 dataset. 

\begin{figure}[ht]
\centering
\begin{subfigure}{0.48\textwidth}
\includegraphics[width=\textwidth]{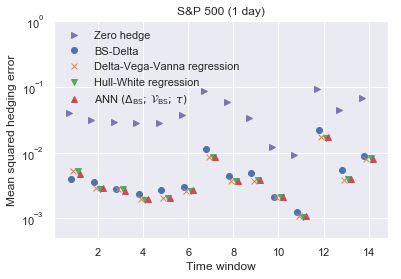}
\end{subfigure}
\quad
\begin{subfigure}{0.48\textwidth}
	\includegraphics[width=\textwidth]{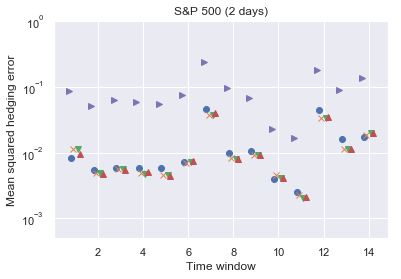}
\end{subfigure}
\caption{MSHEs for four different statistical models of the hedging ratio and the zero hedge across all 14 time windows in the S\&P~500 dataset for the one-day (left) and two-day (right) hedging periods. The numbers of the statistical models correspond to the numbers in \Cref{fig:sp-performance}, but are now presented on a logarithmic scale.
\label{fig:sp-performance-no-hedge}}
\end{figure}

\begin{figure}[ht]
	\centering
	\begin{subfigure}{0.48\textwidth}
	\includegraphics[width=\textwidth]{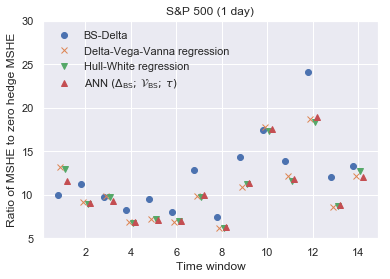}
	\end{subfigure}
	\quad
		\begin{subfigure}{0.48\textwidth}
		\includegraphics[width=\textwidth]{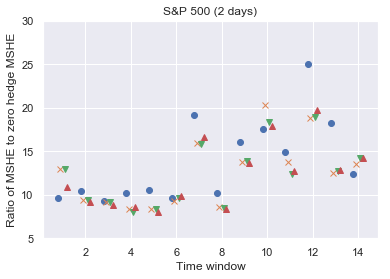}
	\end{subfigure}
	\caption{The ratio of the MSHEs of four statistical models to the hedging ratio and the zero hedge MSHE in the S\&P~500 dataset for the one-day (left) and two-day (right) hedging.}
	\label{fig:sp-variance-ratio-to-nohedge}
\end{figure}

\Cref{fig:sp-performance-no-hedge} extends \Cref{fig:sp-performance} by including the MSHE of the zero hedge strategy.  As we can see, the MSHE for any of the methods is large exactly when the MSHE of the unhedged portfolio is large.  \Cref{fig:sp-variance-ratio-to-nohedge} shows the ratio of the MSHEs of the same four statistical models to the zero hedge MSHE.  The hedging performance gets worse in later periods. The MSHE corresponding to the BS Delta  minus the MSHE of one of the statistical models  divided by the zero hedge MSHE  is about 2\%.

\begin{figure}[ht]
	\centering
		\begin{subfigure}{0.48\textwidth}
	\includegraphics[width=\textwidth]{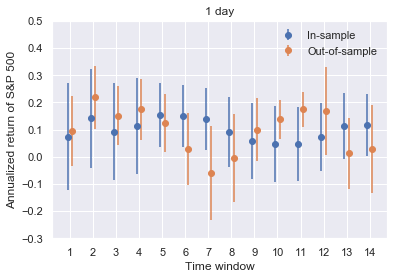}
	\end{subfigure}
	\quad
	\begin{subfigure}{0.48\textwidth}
	\includegraphics[width=\textwidth]{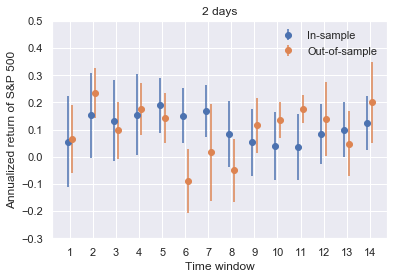}
	\end{subfigure}
	\caption{The average annualised logarithmic one-day (left) and two-day (right) return of the S\&P~500 in each of the 14 time windows. Each line segment shows the average logarithmic return plus/minus one standard error of the logarithmic returns for each time window. The lines tend to be longer, meaning a higher standard deviation, when the returns are smaller, illustrating the leverage effect. }
	\label{fig:sp-return-volatility}
\end{figure}

		\Cref{fig:sp-return-volatility} shows the average  logarithmic return and its standard deviation of the S\&P~500 dataset in each time window. We see that the standard deviations in the out-of-sample sets tend to be large when  the zero hedge MSHEs in \Cref{fig:sp-performance-no-hedge} are large.			

\begin{figure}[ht]
	\includegraphics[width=\textwidth]{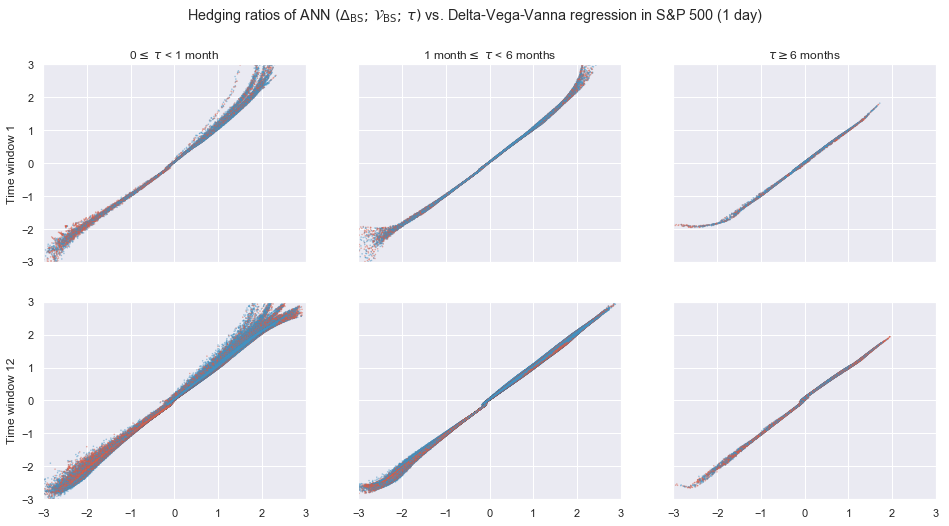}
\caption{ $\text{ANN}(\Delta_{\rm BS};\, \mathcal{V}_{\rm BS};\, \tau)$ versus Delta-Vega-Vanna regression hedging ratios in the S\&P~500 dataset. Each point represents a sample.  We use transformed scales so that the $x$-value of each sample corresponds to $\mathbf{N}^{-1}(\delta_{\rm LR})$ and the  $y$-value  to $\mathbf{N}^{-1}(\delta_{\rm NN})$, where
 $\mathbf N$ denotes the cumulative standard normal distribution. If the point is blue the MSHE corresponding to the ANN is smaller than the one corresponding to the linear regression. On the other hand, if the point is red the linear regression outperforms. 
 Each row shows a time window; the one on the top is a window when the BS Delta outperforms  the statistical models;  the one on the bottom is a more typical one, when the linear regressions and ANNs outperform the BS Delta. Each column corresponds to a different set of maturities; namely less than one month (left), 1 month to 6 months (middle), and more than 6 months (right).
  }
\label{fig:sp-d1-plot}
\end{figure}

\Cref{fig:sp-d1-plot} scatterplots the hedging ratios corresponding to the different statistical models. Here, we provide only one such plot, namely comparing the Delta-Vega-Vanna hedging ratio with the hedging ratio of the $\text{ANN}(\Delta_{\rm BS};\, \mathcal{V}_{\rm BS};\, \tau)$ for the one-day hedging period.  Each point is a sample in the test set. We do not directly plot the hedging ratios but $\mathbf{N}^{-1}(\delta_{\rm NN})$ against $\mathbf{N}^{-1}(\delta_{\rm LR})$, where  $\mathbf N$ denotes again the cumulative standard normal distribution.  The ratios  are very similar but different in the tails, where the ANN seems to overfit.
We only provide the plots for two representative time  windows. In window 1, the BS Delta outperforms all regression models, while window 12 represents a more typical situation where the BS Delta underperforms the regression model and the ANN.

\begin{figure}[ht]
	\centering
	\begin{subfigure}{0.48\textwidth}
		\includegraphics[width=\textwidth]{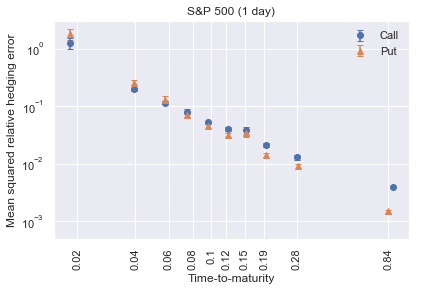}
	\end{subfigure}
	\quad
	\begin{subfigure}{0.48\textwidth}
		\includegraphics[width=\textwidth]{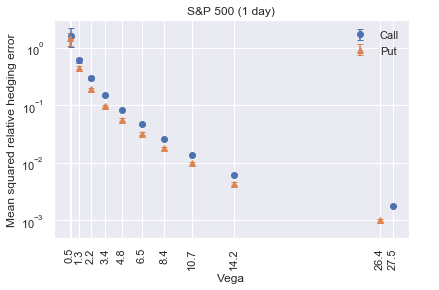}
	\end{subfigure}
\caption{Mean squared relative hedging error of the Delta-Vega-Vanna regression on a logarithmic scale against time-to-maturity (left) and Vega (right) in the S\&P~500 dataset for the one-day hedging period. 
Each line segment provides a point estimate plus/minus one standard error.
Each interval has 10\% of the overall samples, and the tick on the $x$-axis shows the average time-to-maturity and Vega, respectively, of the samples falling into the corresponding interval. Calls and puts may have different averages in each interval. }
\label{fig:sp-pnl-tau-vega}
\end{figure}

\Cref{fig:sp-pnl-tau-vega} plots the mean squared relative hedging error, i.e., the average of the hedging errors divided by the option prices, of the Delta-Vega-Vanna regression against time-to-maturity and Vega. The left panel shows an exponential decrease (due to the logarithmic scale)  of the relative hedging error with respect to time to maturity. The right panel shows that the relative hedging errors decrease super-exponentially as Vega increases, i.e., as the options have a longer time-to-maturity and are less out-of-the money.

\section{Additional diagnostics for  the Euro Stoxx~50 dataset} \label{A:addition Eurox50}
Similar to Appendix~\ref{A:addition SP500} we now provide some additional figures for the Euro Stoxx~50 dataset.

\begin{figure}[ht]
	\includegraphics[width=\textwidth]{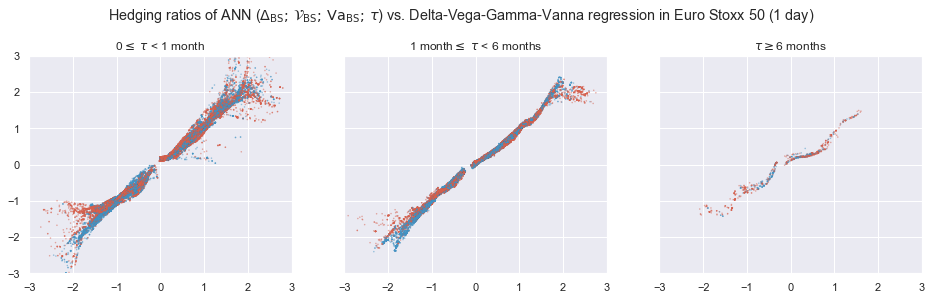}
	\caption{$\text{ANN}(\Delta_{\rm BS};\, \mathcal{V}_{\rm BS};\, {\rm Va}_{\rm BS};\, \tau)$ versus Delta-Vega-Gamma-Vanna regression hedging ratios in the Euro Stoxx~50 dataset. See \Cref{fig:sp-d1-plot} for additional explanations. 
	}
	\label{fig:st-d1-plot}
\end{figure}

\Cref{fig:st-d1-plot} scatterplots the hedging ratios corresponding to the different statistical models.  We refer to the caption of \Cref{fig:sp-d1-plot} for explanations.  Different to \Cref{fig:sp-d1-plot} with the S\&P~500 dataset, the hedging ratios of the ANN now look quite different from the linear regression model. Consistently with the prevalence of red points, for the one-day hedging period, the ANNs display a relatively bad performance (recall \Cref{tab:st-overall}).

\begin{figure}[ht]
	\centering
	\begin{subfigure}{0.48\textwidth}
		\includegraphics[width=\textwidth]{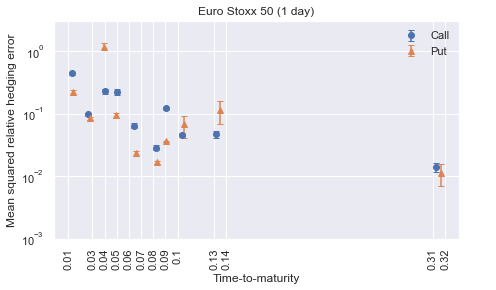}
	\end{subfigure}
	\quad
	\begin{subfigure}{0.48\textwidth}
		\includegraphics[width=\textwidth]{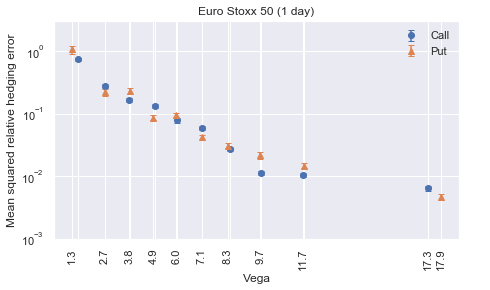}
	\end{subfigure}
\caption{Mean squared relative hedging error of the Delta-Vega-Gamma-Vanna regression on a logarithmic scale against time-to-maturity (left) and Vega (right) in the Euro Stoxx~50 dataset  for the one-day hedging period.  See \Cref{fig:sp-pnl-tau-vega} for additional explanations. }	
\label{fig:st-pnl-tau-vega}
\end{figure}

\Cref{fig:st-pnl-tau-vega} plots the mean squared relative hedging error, i.e., the average of the hedging errors divided by the option prices, of the Delta-Vega-Gamma-Vanna regression against time-to-maturity and Vega. In comparison to the S\&P~500 dataset (see \Cref{fig:sp-pnl-tau-vega}), the decrease seems to be a little bit smaller as time-to-maturity and Vega increase, respectively.

\end{appendices}

\end{document}